\documentclass[aps,prresearch,reprint,,tightenlines,superscriptaddress]{revtex4-1}

\usepackage{amsmath,amsfonts,amssymb}
\usepackage{mathrsfs}
\usepackage{xparse}
\usepackage{physics}
\usepackage{booktabs}
\usepackage{graphicx}
\usepackage[utf8]{inputenc}
\usepackage[english]{babel}
\usepackage{hyperref}
\usepackage{mhchem}

\renewcommand{\mathbf}{\boldsymbol}
\DeclareDocumentCommand\term{mmg}{%
  {$^{#1}$#2%
  \IfNoValueF {#3} {$_{#3}$}%
  }%
}

\makeatletter
\newcommand\etal{\emph{et al}\@ifnextchar.{}{.\ }}
\newcommand\eryso{Er$^{3+}$:Y$_2$SiO$_5$\@ifnextchar.{}{\@ifnextchar,{}{ }}}
\newcommand\yso{Y$_2$SiO$_5$\@ifnextchar.{}{\@ifnextchar,{}{ }}}
\newcommand\peryso{$^{167}$Er$^{3+}$:Y$_2$SiO$_5$\@ifnextchar.{}{\@ifnextchar,{}{ }}}
\makeatother

\begin{document}
\title{Noise Free On-Demand Atomic Frequency Comb Quantum Memory}

\author{Sebastian P.\ Horvath}
\thanks{These authors contributed equally.}

\author{Mohammed K.\ Alqedra}
\thanks{These authors contributed equally.}

\author{Adam Kinos}
\author{Andreas Walther}
\author{Jan Marcus Dahlstr\"om}
\author{Stefan Kr\"oll}
\author{Lars Rippe}
\email{physics@rippe.se}
\affiliation{Department of Physics, Lund University, P.O. Box 118, SE-22100 Lund, Sweden.}

\date{\today}


\begin{abstract}
We present an extension of the atomic frequency comb protocol that utilizes the Stark effect to perform noise-free, on-demand, control. An experimental realization of this protocol was implemented in the Pr$^{3+}$:Y$_2$SiO$_5$ solid-state system, and a recall efficiency of 38\% for a 0.8 $\mu$s storage time was achieved. Experiments were performed with both bright pulses as well as weak-coherent states, the latter achieving a signal-to-noise ratio of $570 \pm 120$ using input pulses with an average photon number of $\sim 0.1$. The principal limitation for a longer storage time was found to be the minimum peak width attainable for Pr$^{3+}$:Y$_2$SiO$_5$. We employ an adaptation of an established atomic-frequency comb model to investigate an on-demand, wide-bandwidth, memory based on Eu$^{3+}$:Y$_2$SiO$_5$. From this we determine that a storage time as long as 100 $\mu$s may be practical even without recourse to spin-wave storage. 

\end{abstract}

\maketitle

\section{INTRODUCTION}

Geographically distributed quantum entanglement is a fundamental resource in quantum cryptography and quantum-information science. Optical fiber losses necessitate quantum repeaters for distances exceeding a few tens of kilometers, an integral part of which are optical quantum memories. In particular, quantum memories with a storage time of at least $\sim 100$ $\mu$s are required for second generation quantum repeaters \cite{muralidharan_2016}. A range of schemes have been proposed for realizing such a memory, including electromagnetically induced transparency, in both atomic vapors as well as solid-state media \cite{longdell_2005,hsu_2006,schraft_2016}, Raman transfer memories \cite{reim_2011, sprague_2014}, gradient echo memories/controlled-reversible inhomogeneous broadening (GEM/CRIB) \cite{nilsson_2005,hetet_2008,hosseini_2011,hedges_2010,cho_2016}, revival of silenced echo (ROSE) \cite{Dajczgewand_ROSE_2014}, spectral hole memory \cite{Kutluer_Holes_2016}, hybrid photon-echo rephasing (HYPER) \cite{McAuslan2011_HYPER}, and atomic frequency combs (AFC) \cite{afzelius_multimode_2009,minar_2009,afzelius_impedance_2010,sabooni_2013,jobez_2015,gundogan2015}. The principal figures of merit are the recall efficiency and the storage time, as well as the bandwidth and the noise performance. To date, finding a memory protocol that can satisfy all of these requirements simultaneously remains an outstanding problem.  

In this article, we demonstrate an on-demand AFC protocol with background-free readout at the single photon level. An AFC memory consists of an ensemble of highly absorbing ions that are periodic in frequency. Storage light incident on the comb results in a collective excitation, which dephases because of the energy detuning between the absorbing peaks. The periodicity of the ions leads the collective excitation to be in phase again after a time corresponding to the inverse of the comb spacing, thus yielding a coherent echo. A principal challenge with the standard AFC protocol is that, in order to be on-demand, it must be combined with a transfer of the excitation to the ground-state hyperfine levels. While this can additionally enhance the storage time, the coherent transfers require bright optical pulses, which necessitate complex filtering to separate the signal photons from coherent background emission \cite{timoney_single_2013,bonarota_2014,jobez_2015,gundogan2015}. Here we demonstrate a modification that utilizes the linear Stark effect to divide the atomic ensemble into two spectroscopic ion classes and then switches these two classes out of phase by applying an electric-field pulse. Thus, it is possible to quench the echo emission in all direction until a second, equivalent, electric field pulse is applied. 
This technique has previously been used in connection with optically detected spin echo experiments \cite{arcangeli_2016}. In the present work, this electric field control approach is used in conjunction with an AFC memory, creating an on-demand variant of the AFC protocol which does not rely on optical-control
pulses and is consequently noise free. We observe that, while the experimental work was performed using weak-coherent states, the linearity of this memory scheme means it will work equally well for discrete and continuous variable storage \cite{hedges_2010,lobino_2009}. 

\begin{figure*}[tb!]
\centering
\includegraphics[width=\textwidth]{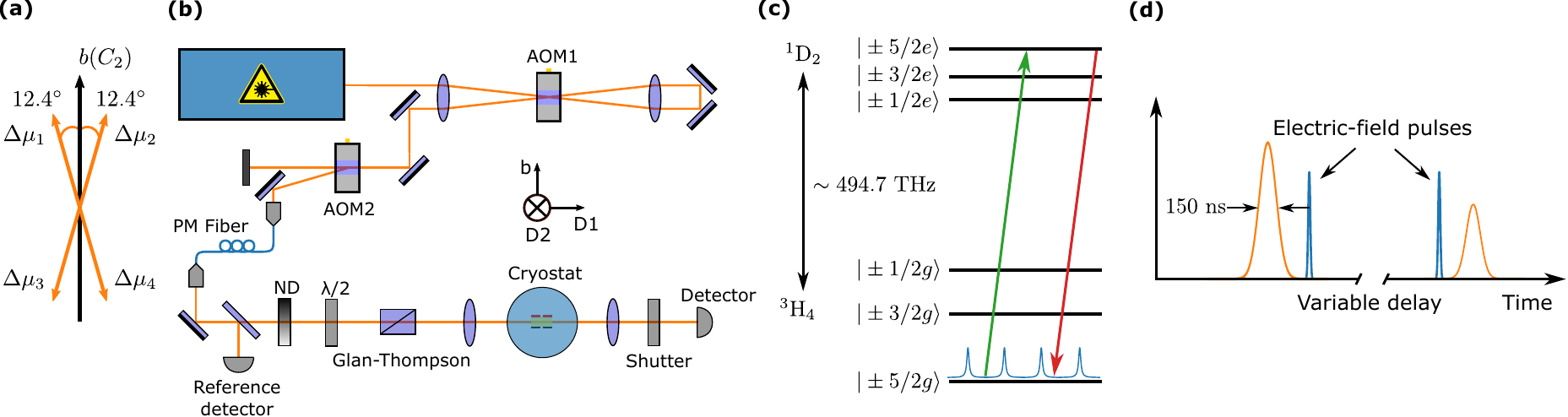}
\caption{(a) The orientation of the permanent electric dipole moment differences between the $^3$H$_4$ ground and $^1$D$_2$ excited state in Pr$^{3+}$:Y$_2$SiO$_5$. They all lie at an angle of 12.4$^{\circ}$ with respect to the $b$ axis in a plane tilted 35$^{\circ}$ from the $D_2$ axis \cite{equall_ultraslow_1994,graf_1997,beavan_2013}. (b) The experimental setup. The light is polarized with the electric field parallel to the crystallographic $D_2$ axis. (c) The energy-level schematic of Pr$^{3+}$:Y$_2$SiO$_5$. The primary contribution to the AFC structure was due to four distinct groups of ions resonant with the $\ket{\pm5/2g} \to \ket{\pm 5/2e}$ transition. (d) The implemented pulse sequence, showing the 1.2 $\mu$s input pulse, the electric-field pulses, and the AFC echo. The delay between the electric-field pulses was adjusted to obtain different storage times.}
\label{fig:exp}
\end{figure*}

The AFC protocol has earlier been combined with CRIB to yield an electrically controlled variant of the AFC memory, where a spatially varying electric field prevents the build up of echo emission in the phase matching direction \cite{lauritzen_2011}. However, the recall efficiency of higher-order echoes proved to be low, possibly due to poor ensemble preparation. A recall up to the second order echo was shown and the ratio compared to no field was $<$1. Here we show a recall up to the tenth echo with $>$10 ratio compared to no field. The CRIB based switching of the AFC is distinct from our switching in that the electric-field gradient dephases atoms on a macroscopic scale along the direction of propagation. In contrast, our protocol operates by switching groups of atoms out of phase on the sub-micron scale by utilizing homogeneous, discrete, electric field pulses.
Furthermore, the presented scheme has the advantage of significantly minimizing the delay between the retrieval decision and the retrieval occurrence of the stored photon. This delay is mainly limited by the Stark pulse duration, which can potentially be as short as a few nanoseconds, and the duration of the stored photon. Consequently, very high recall resolution with minimal delay between the retrieval decision and the re-emission can be achieved by choosing an appropriate comb spacing and using short Stark pulses. Although the technique in Ref \cite{lauritzen_2011} is different from that in Ref \cite{arcangeli_2016} and in this paper, it is not unreasonable to expect it could perform similarly.

\section{THEORY}

Using the linear Stark effect for phase control in rare-earth ion-doped materials was originally demonstrated for photon-echo intensity modulation \cite{wang_1992,meixner_1992,graf_1997}. We will discuss the concept for applications using a yttrium orthosilicate (Y$_2$SiO$_5$) host crystal. For a dopant ion substituted at the yttrium site there are four classes of ions with distinct permanent electric dipole moment orientations; Fig. \ref{fig:exp}(a) shows the orientations for Pr$^{3+}$:Y$_2$SiO$_5$.

Due to the space-group symmetry of Y$_2$SiO$_5$, an electric field applied along the crystallographic $b$ axis leads to two distinct ion classes that experience a Stark shift of equal magnitude but with opposite sign. These will henceforth be referred to as the positive and negative electrically inequivalent ion classes. The Stark shift magnitude is given by $\Omega = [(\mathbf{\mu}_g - \mathbf{\mu}_e)/h] \cdot \mathbf{E}$, for ground- and excited-state permanent electric dipole moments $\mathbf{\mu}_g$ and $\mathbf{\mu}_e$, an electric field $\mathbf{E}$, and Planck constant $h$. Consequently, if an electric-field pulse is applied, the ions are frequency shifted and accumulate a relative phase with respect to their unperturbed state.
Neglecting spontaneous decay and homogeneous dephasing, after absorbing a photon at time $t=0$, the state of an AFC can be described as a collective excitation of the form
\begin{equation}
	\ket{\psi(t)} = \frac{1}{\sqrt{2M}}\sum_{\ell = 0}^{M-1} e^{i \omega_{\ell} t} \left[e^{i 2\pi \Omega T} \ket{\psi_{\ell}^+} + e^{-i 2\pi \Omega T} \ket{\psi_{\ell}^-}\right].
	\label{eqn:col_ex}
\end{equation}
Here, $M$ is the number of AFC peaks, $\omega_{\ell} = 2 \pi \Delta \ell$, where $\Delta$ is the separation between the AFC peaks, and $T$ is the length of the electric-field pulse. Furthermore, $\ket{\psi_{\ell}^+}$ and $\ket{\psi_{\ell}^-}$ are the wavefunctions of the positive and negative electrically inequivalent ion classes, with the former defined as
\begin{equation}
	\ket{\psi_{\ell}^+} = \frac{1}{\sqrt{N_{\ell}^+}}\sum_{j = 1}^{N_{\ell}^+} c_{\ell j}^+ e^{2 \pi i \delta_{\ell j}^+ t} e^{-i k z_{\ell j}^+} \ket{g_1 \ldots e_j \ldots g_{N_{\ell}^+}},
	\label{eqn:psi_plus}
\end{equation}
where $N_{\ell}^+$ is the number of ions in peak $\ell$ that experience a positive frequency shift because of $\mathbf{E}$, $c_{\ell j}^+$ is the probability amplitude that the input photon excites ion $j$ in AFC peak $\ell$, $\delta_{\ell j}^+$ is the frequency detuning of the $j$th ion from the AFC peak center frequency $\omega_\ell$, $k$ is the photon wave vector, and $z_{\ell j}^+$ is the position of ion $j$ in the AFC peak $\ell$ along the propagation direction. $\psi_{\ell}^-$ is described by an analogous expression to Eq. \eqref{eqn:psi_plus}. We note that normalizing $\ket{\psi(t)}$ leaves the individual wavefunctions $\ket{\psi_{\ell}^+}$ and $\ket{\psi_{\ell}^-}$ unnormalized. Additionally, it is assumed that for each $\ell$ the distributions $\delta_{\ell j}^{\pm}$ are approximately Gaussian with a full-width at half maximum (FWHM) $\gamma = 140$ kHz, and $>10^{10}$ ions per peak. Moreover, we consider an equal probability of occupation of the electrically inequivalent sites, leading to $\vert N_{\ell}^+ - N_{\ell}^- \vert \approx \sqrt{N_{\ell}^+} \approx 10^5$.

\begin{figure*}[tbh!]
\centering
\includegraphics[width=\textwidth]{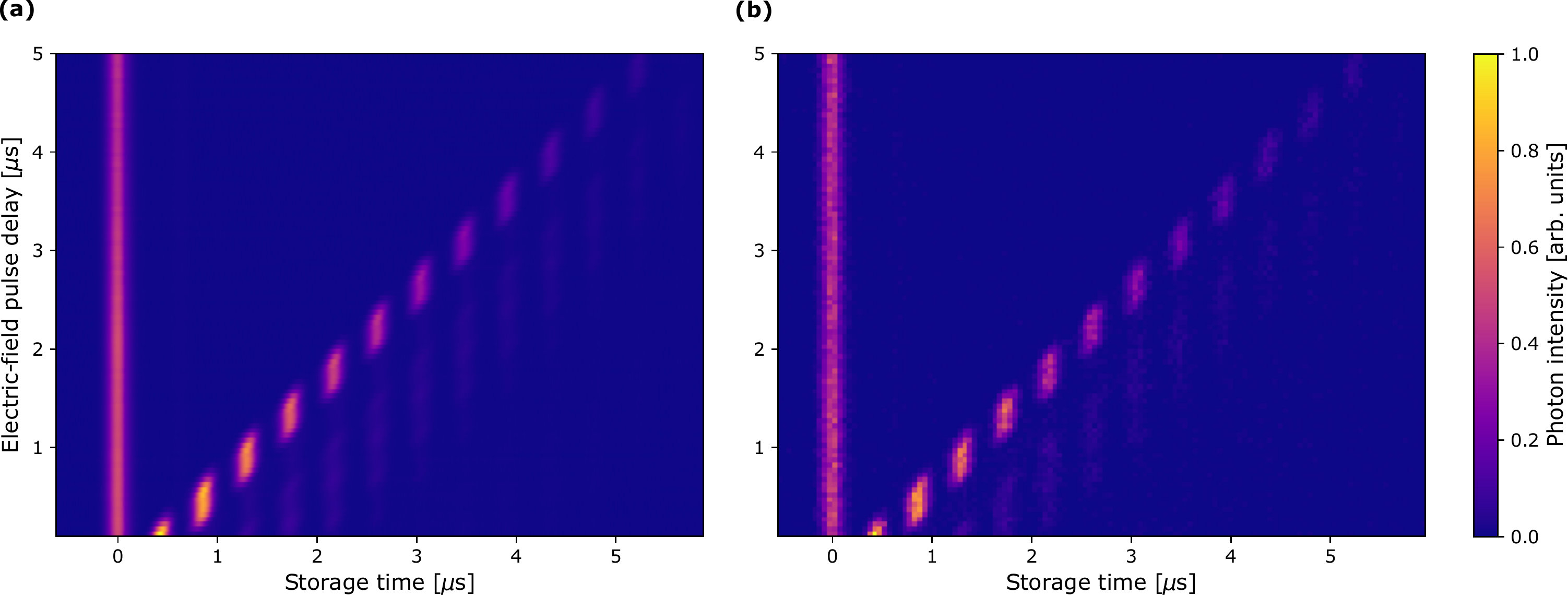}
\caption{(a) The photon transmission through the AFC memory while altering the delay between the two electric-field pulses to inhibit and enable echo emission, using bright optical input pulses. The AFC echo emission is discrete in time and occurs at times corresponding to an integer multiple of $1/\Delta \approx 430$ ns. The light at $\sim 1$ $\mu$s corresponds to photons not absorbed by the memory. (b) An equivalent set of measurements to (a) using weak-coherent states. Storage pulses were attenuated to $0.097 \pm 0.004$ photons per shot. Each ``Electric-field pulse delay'' slice corresponds to an accumulation of $3 \times 10^4$ shots.}
\label{fig:echo_maps}
\end{figure*}

From Eq. \eqref{eqn:col_ex} it can be seen that at $t=0$ all excited ions will oscillate in phase; however, after a short time $\delta t$ the accumulated phase between peaks will lead to destructive interference. Considering the scenario where no external electric field is applied ($T = 0$), all excited ions will again oscillate in phase at times $t = m/\Delta$ for $m \in \mathbb{Z}$ (provided $m/\Delta \ll \gamma$, the peak width). In contrast, if an electric-field pulse with $T = 1/(4 \Omega)$ is applied the contribution from the two wave functions $\ket{\psi_{\ell}^+}$ and $\ket{\psi_{\ell}^-}$ will have opposite phase and the photon emission probability will be reduced by approximately ten orders of magnitude. Thus, with such an electric-field pulse the AFC photon emission can, effectively, be completely inhibited. If, however, a second equivalent electric-field pulse is applied, for example in the time interval $(n-1)/\Delta < t < n/\Delta$, the contributions of $\ket{\psi_{\ell}^+}$ and $\ket{\psi_{\ell}^-}$ will be in phase and lead to photon emission at $t=n/\Delta$, for $n \in \mathbb{Z}$. Consequently, by applying one electric-field pulse during the time interval $[0, 1/\Delta]$, and a second electric-field pulse in the time interval $[(n-1)/\Delta, n/\Delta]$, one obtains on demand photon emission at $t=n/\Delta$. 

\section{METHOD}

The experiment employed a Y$_2$SiO$_5$ crystal (Scientific Materials Inc.) with Pr$^{3+}$ ions substituted for Y$^{3+}$ at a 500 ppm level, cooled to $2.1$ K. The sample had dimensions of $6$ by $10$ by $10$ mm, with respect to the crystallographic $b$ axis, and the optical extinction axes $D_1$ and $D_2$. The two faces normal to the $b$ axis were coated by a pair of gold electrodes, which could be used to apply an electric field in the $b$ direction. Further details on the sample can be found in Ref.\ \cite{li_2016}.

The light source was a Coherent 699-21 ring dye laser, locked to an external ULE cavity at 494.723 THz, on resonance with the $^3$H$_4$ $\to$ $^1$D$_2$ transition of Pr$^{3+}$. The experimental arrangement is shown in Fig.\ \ref{fig:exp}(b). Light pulses could be arbitrarily tailored in phase, frequency, and amplitude with two acousto-optic modulators in series; optional further attenuation was provided by a neutral density filter attached to a motorized mount. For weak-coherent state measurements, photon detection was performed with a Laser Components Count 50N avalanche photodiode with a quantum efficiency of 0.69 at 606 nm and a dark-count rate of 26 Hz.
 
To prepare the AFC, an 18 MHz transmission window was burned at the center of the inhomogeneous line following the procedure of Ref.\ \cite{amari_2010}. The comb structure consisted of four distinct groups of ions, separated by intervals of $\Delta = 2.3$ MHz for their $\ket{5/2g} \to \ket{5/2e}$ transition, stored in the $\ket{5/2g}$ state. The comb contained two further ion groups that had either the $\ket{3/2g} \to \ket{1/2e}$ or the $\ket{3/2g} \to \ket{3/2e}$ transition frequencies overlapped with the above indicated $\ket{5/2g} \to \ket{5/2e}$ transition frequencies. The dominant $\ket{5/2g} \to \ket{5/2e}$ transition is shown in Fig.\ \ref{fig:exp}(c). Further details on the AFC creation, as well as absorption spectra of the comb structure, are provided in Appendix \ref{sec:afc_burn}.

\section{RESULTS AND DISCUSSION}

The storage light consisted of a Gaussian pulse of FWHM $=150$ ns resonant with the center of the comb structure. Prior to emission of the echo, an electric-field pulse of magnitude $54$ V and FWHM $=23$ ns was applied to the gold-coated electrodes, which corresponded to a relative phase shift of $\pm \pi/2$ for the two electrically inequivalent ion classes, thus quenching the coherent emission. By waiting a variable amount of time prior to applying a second, equivalent, electric-field pulse, the recall efficiency with respect to time was mapped out. The corresponding pulse sequence is shown in Fig.\ \ref{fig:exp}(d). In Figs. \ref{fig:echo_maps}(a) and \ref{fig:echo_maps}(b) the efficiency maps are shown for bright storage pulses and weak-coherent states. The weak-coherent state input power calibration was performed by removing the crystal from the beam path and adjusting the attenuation to achieve a count rate of $0.097 \pm 0.004$ photons per storage pulse, compensating for the detector quantum efficiency and cryostat window losses. We performed 2000 storage-pulse shots per AFC creation cycle and accumulated photon counts over 15 AFC creation cycles per ``Electric-field pulse delay'' slice in Fig.\ \ref{fig:echo_maps}(b).

Figure \ref{fig:echo_maps}(a) shows that switching the two ion classes out of phase quenches the emission and there is virtually no residual light after the application of a single $\pm \pi/2$ pulse. From the weak-coherent state measurement of Fig.\ \ref{fig:echo_maps}(b) the advantage of performing the recall control without optical pulses is apparent. In particular, we obtain a signal-to-noise ratio (SNR) of $570 \pm 120$ for a storage time of 860 ns using input pulses with an average photon number of $0.097 \pm 0.004$. The SNR is calculated by dividing the total number of signal photons in a $350$ ns time bin, accumulated over $3 \times 10^4$ consecutive storage shots, by the noise floor. Here, the noise floor is defined as the number of dark counts in a time bin of equivalent length, also accumulated over $3 \times 10^4$ consecutive shots. Thus, for single-photon Fock states we expect the SNR to improve by an order of magnitude. Furthermore, the unconditional noise floor \cite{reim_2011,timoney_single_2013} is effectively the dark-count rate of the detection setup, since the optical lifetime of site 1 in Pr$^{3+}$:Y$_2$SiO$_5$ is 164 $\mu$s \cite{equall_1995} and we utilized a 200 ms waiting time between the comb creation and the storage pulses. To put this SNR into perspective, Gündoğan \textit{et al} \cite{gundogan2015} estimated an SNR of $\sim 20$ for spin-storage time of 1 $\mu$s using input pulses with one photon per pulse on average.

Due to the Stark-shift modulated phase control, the AFC storage time is not limited by the comb rephasing time $1/\Delta$. Nevertheless, as is apparent from Figs.\ \ref{fig:echo_maps}(a) and (b), the recall efficiency drops considerably for longer storage times. This is a consequence of the comb-peak width, which leads to a decoherence of the peak ensembles. Modifying the analytical model for Gaussian AFC peaks of Ref.\ \cite{afzelius_multimode_2009} to account for the Stark-shift phase control, one obtains the following efficiency expression
\begin{equation}
	\eta = (\tilde{\alpha} L)^2 e^{-\tilde{\alpha} L} e^{-t^2 \tilde{\gamma}^2},
	\label{eqn:eta}
\end{equation}
where $t$ is the storage time. Further, $\tilde{\gamma}$ is related to the comb FWHM by $\tilde{\gamma} = \frac{2 \pi \gamma}{\sqrt{8 \ln(2)}}$, $\tilde{\alpha}$ is the effective absorption of the comb defined as $\tilde{\alpha} = \frac{\alpha}{F}\sqrt{\frac{\pi}{4 \ln(2)}}$, with $F$ the comb finesse, and $\alpha$ the absorption of the comb peaks.

\begin{figure}[h]
\centering
\includegraphics[width=1.05\columnwidth]{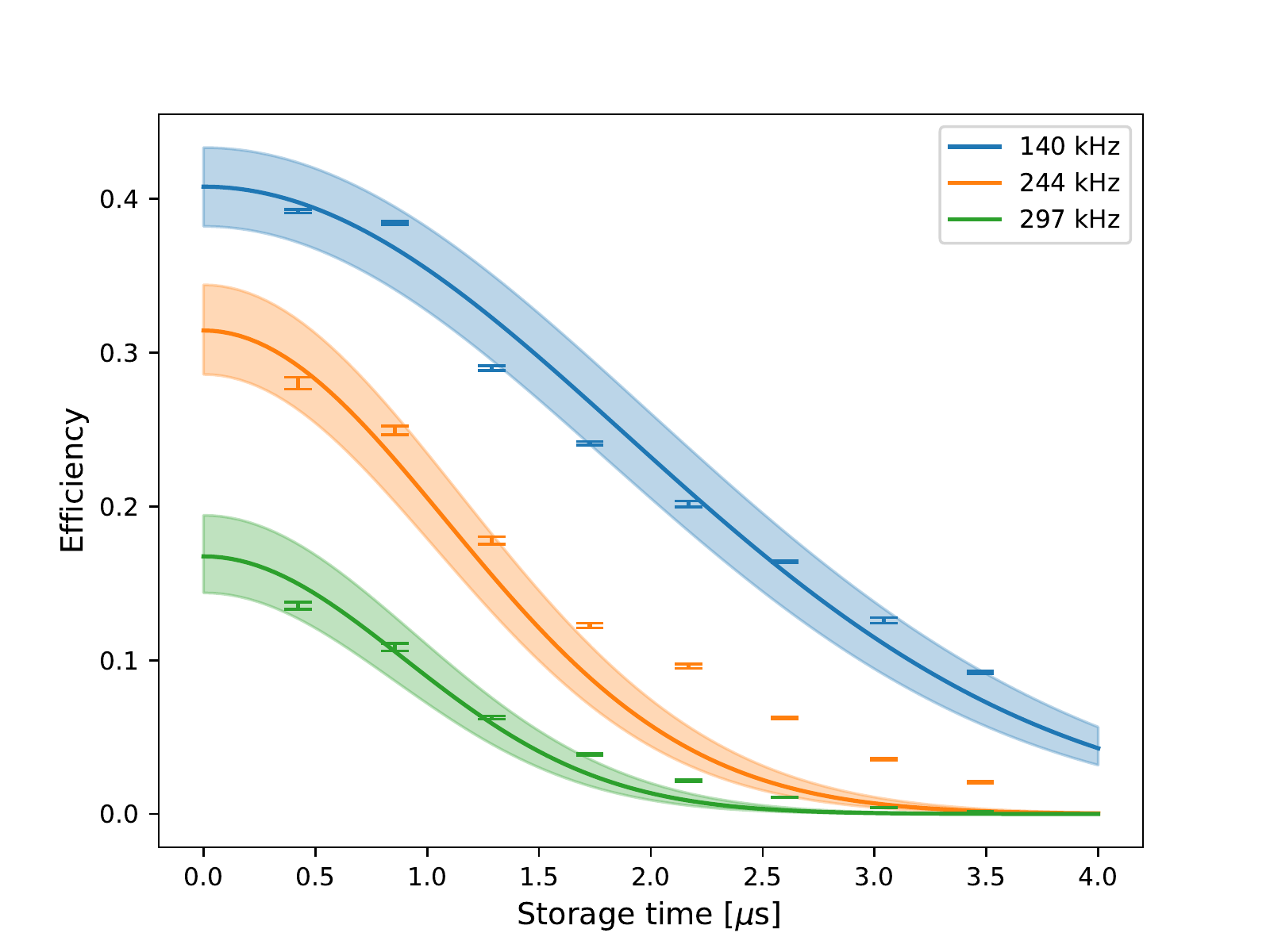}
\caption{Recall efficiency over a range of storage times for several AFC peak widths. The solid line represents the model fit, with the shaded region indicating the uncertainty due to the peak width, which was estimated to be $\pm 5$\% from several absorption measurements.}
\label{fig:pw_dep}
\end{figure}

Figure \ref{fig:pw_dep} shows the efficiency with respect to readout time for three different peak widths. As can be seen in the figure, the efficiency decay rate is only limited by the AFC peak width, which indicates that our scheme is phase preserving and capable of storing quantum states with high fidelity. The efficiency was estimated by comparing the recall intensity of the first echo with a pulse of equal magnitude propagated through the 18 MHz spectral window without any absorbing structure. Such a transmission measurement was previously verified to be within a few percent of the transmission through our optical setup with the crystal removed. The solid lines correspond to Eq.\ \eqref{eqn:eta}. Because of some uncertainty in the peak heights, the $\alpha$ value for each peak width was varied to achieve a good fit. Additionally, we note that the comb creation for the 244 kHz and 297 kHz peak widths yielded some deviation from Gaussian lineshapes. Further details of the peak characterization, as well as data for additional peak widths, can be found in Appendixes \ref{sec:readout} and \ref{sec:peak width}.

Finally, we investigate the memory efficiency that could, in principle, be obtained using the presented protocol. For an implementation using Pr$^{3+}$:Y$_2$SiO$_5$ the primary limitation to a long storage time is the minimum peak width, which is dictated by the optical coherence time of 152 $\mu$s \cite{equall_1995}. Equation \eqref{eqn:eta} allows us to benchmark materials with longer coherence times. Furthermore, we note that for an AFC memory with an optical cavity, the dephasing term is identical to the last exponential in Eq.\ \eqref{eqn:eta}.  Thus, the efficiency for a Stark modulated AFC memory with an optical cavity reads
\begin{equation}
	\eta_{\text{cav}} = \frac{4 (\tilde{\alpha} L)^2 e^{-2 \tilde{\alpha} L} (1-R_1)^2 R_2 e^{-t^2 \tilde{\gamma}^2}}{(1-\sqrt{R_1 R_2} e^{-\tilde{\alpha} L})^4},
\end{equation}
where $R_1$ and $R_2$ are the mirror reflectivities, assuming no scattering losses \cite{afzelius_impedance_2010}. Evaluating the efficiency for a comb linewidth of 1 kHz, and an $\alpha L = 1$ \cite{konz2003}, we obtained an efficiency of $88\%$ for a storage time of 100 $\mu$s, with an impedance matched cavity employing mirrors with reflectivities of $R_1 = 96\%$ and $R_2 = 99.9\%$. Such parameters may be achievable using a Eu$^{3+}$:Y$_2$SiO$_5$ crystal for which persistent holes with a width in the order of kHz have been used extensively for laser frequency stabilization \cite{thorpe2011,leibrandt2013,gobron2017}. For telecom photons, Eu$^{3+}$:Y$_2$SiO$_5$ memories would require upconversion, e.g. using periodically poled nonlinear crystals. Such a frequency conversion may have an intrinsic efficiency of 100$\%$ for single photons also \cite{Lijun_upconversion}. 
An alternative material, avoiding the complication with frequency conversion, may be Er$^{3+}$:Y$_2$SiO$_5$. 
However, erbium is a Kramer’s ion, and it is generally very difficult to achieve good hole-burning in Kramer’s ions, which is a key requirement for AFC based quantum memories. While Rancic \emph{et al}.\ have demonstrated successful hole-burning in spin-polarized \ce{^{167}Er^{3+}}:Y$_2$SiO$_5$ with a large external magnetic field \cite{Rancic_2018}, the necessary conditions may be challenging to realize simultaneously with the magnetic field requirements necessary for narrow homogeneous linewidths \cite{bottger_2009,bottger_2006}.

\section{CONCLUSIONS}

To conclude, we have introduced an on-demand variant of the AFC quantum memory protocol, which is essentially noise free. We have experimentally demonstrated the operation of such a memory using both bright storage pulses as well as with a weak-coherent state measurement. The weak-coherent state measurement was estimated to achieve an SNR of $570 \pm 120$ using pulses with an average photon number of $0.097 \pm 0.004$. In addition, we have corroborated an analytical model for the memory efficiency with respect to the comb-peak width, and extrapolated this to a narrow linewidth memory based utilizing a cavity. We estimate that an Eu$^{3+}$:Y$_2$SiO$_5$ memory utilizing the Stark modulated AFC protocol in conjunction with an optical cavity could achieve an efficiency of $88\%$ for a storage time of 100 $\mu$s.

\section{ACKNOWLEDGMENTS}

This research was supported by the Swedish Research Council (no. 2016-05121, no. 2015-03989, no. 2016-04375 and 2019-04949), the Knut and Alice Wallenberg Foundation (KAW 2016.0081), the Wallenberg Center for Quantum Technology (WACQT) funded by The Knut and Alice Wallenberg Foundation (KAW 2017.0449), and the European Union FETFLAG program, Grant No. 820391 (SQUARE) (2017.0449).

\appendix
\section{Atomic frequency comb creation \label{sec:afc_burn}}

\begin{figure}[h]
\centering
\includegraphics[width=\columnwidth]{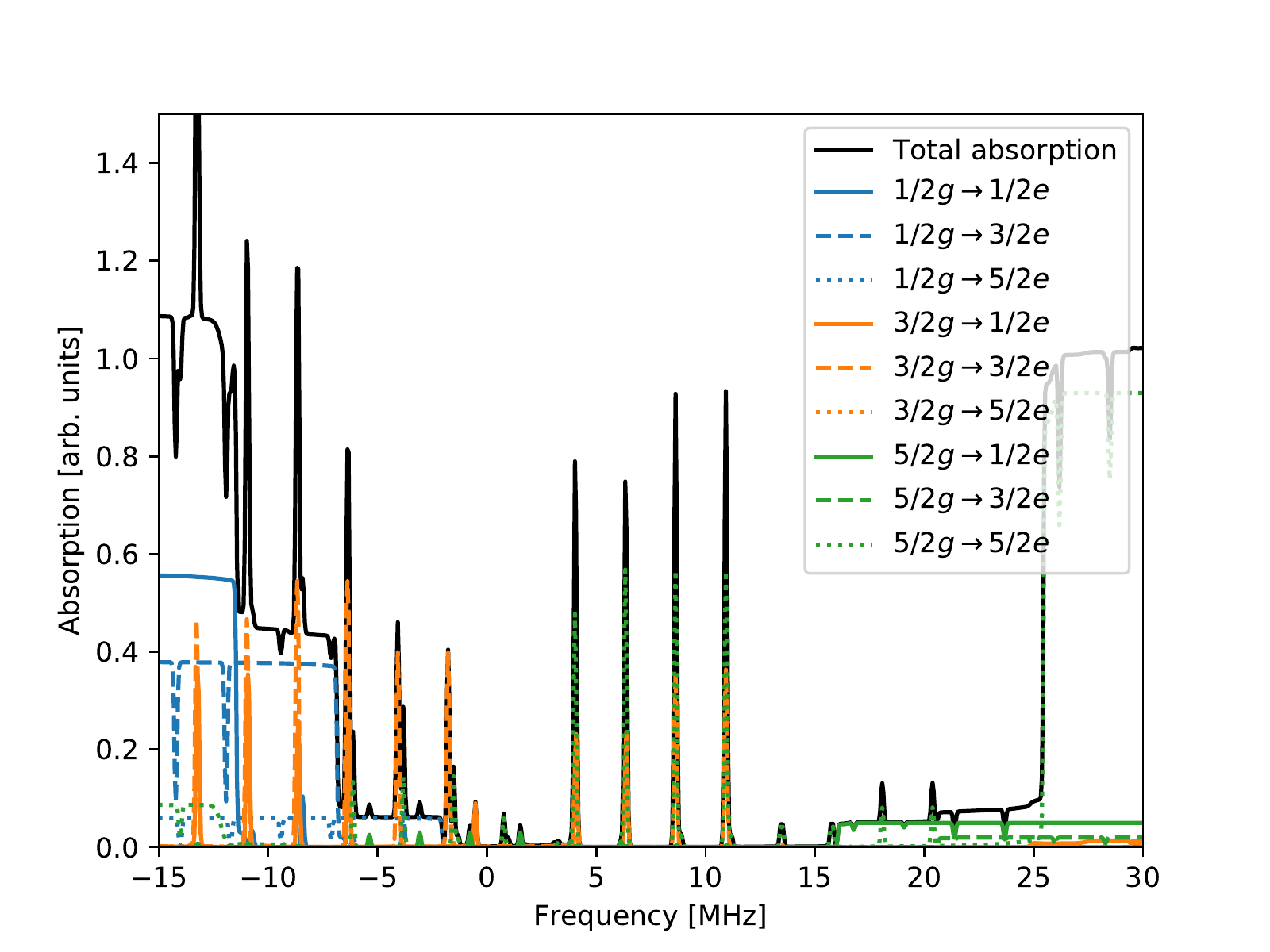}
\caption{The simulated absorption structure obtained from the AFC burn-back sequence described in Tab.\ \ref{tab:BurnbackAFC}. Blue, orange, and green colors correspond to ions in the $\ket{1/2g}$, $\ket{3/2g}$, and $\ket{5/2g}$ hyperfine states. The solid, dashed, and dotted lines correspond to transitions resonant with the $\ket{1/2e}$, $\ket{3/2e}$, and $\ket{5/2e}$ hyperfine states. The AFC peaks are formed by stacking different groups of ions on top of each other, such that the $\ket{3/2g}$ and $\ket{5/2g}$ ions are overlapped in frequency. The black line shows the integrated total absorption.}
\label{fig:afc_struct}
\end{figure}

An 18 MHz wide spectral transmission window was burned into the inhomogeneous absorption profile utilizing a pulse sequence that has been described in detail in Ref.\ \cite{amari_2010}. The pulse sequence to create the atomic frequency comb (AFC) was then optimized by simulation. The simulation kept track of the absorption frequencies of different groups of ions for the hyperfine levels of the ${}^{3}\mathrm{H}_4(1) \to {}^1\mathrm{D}_2(1)$ transition. A hyperbolic tangent transfer efficiency profile of 5\% was assumed for all pulses, which was scaled by the relative dipole moment strength for transitions between the various ground and exited state hyperfine levels. The resulting simulated absorption structure is presented in Fig.\ \ref{fig:afc_struct}, showing the contributions of transitions between individual hyperfine levels as well as the combined absorption profile. 

Table \ref{tab:BurnbackAFC} summarizes the pulse sequence implemented in experiment that was developed using the simulation results. All frequencies are defined with respect to an arbitrary group of ions that has the $\ket{1/2g} \to \ket{1/2e}$ transition frequency coincident with the low-frequency edge of the spectral transmission window, which is set to zero. All transfers were performed using complex hyperbolic secant (sechyp) pulses \cite{rippe_2005,roos_2004}, while class-cleaning pulses utilized a hybrid pulse combining the edges of a sechyp pulse with a linear frequency chirp. 
\begin{table}[!htbp]
\label{tab:BurnbackAFC} 
\centering
\resizebox{\columnwidth}{!}{%
\begin{tabular}{lcccccc}  
 \hline
 \hline
  Pulse name & $\nu$ (MHz) & $\nu_{\mathrm{width}}$ (MHz) & $t_{\mathrm{FWHM}}$ ($\mu$s) & $t_\mathrm{cutoff}$ ($\mu$s) & Target transition & Reps\\
 \hline
  BurnbackAFC1 & 21.34  & 0.1 & 11.2 & 40 & $\ket{5/2g} \to \ket{1/2e}$ & 20\\
  BurnbackAFC2 & 23.64  & 0.1 & 11.2 & 40 & $\ket{5/2g} \to \ket{1/2e}$ & 20\\  
  CleanAFC1    & -1.6   & 0.5 & 3.0  & 86 & $\ket{1/2g} \to \ket{5/2e}$ & 60\\  
  CleanAFC2    & -0.4   & 0.5 & 3.0  & 86 & $\ket{1/2g} \to \ket{5/2e}$ & 20\\ 
  CleanAFC3    &  0.8   & 0.5 & 3.0  & 86 & $\ket{1/2g} \to \ket{5/2e}$ & 60\\ 
  CleanAFC4    &  1.9   & 0.5 & 3.0  & 86 & $\ket{1/2g} \to \ket{5/2e}$ & 60\\
  BurnbackAFC3 & -16.55 & 0.1 & 11.2 & 40 & $\ket{1/2g} \to \ket{5/2e}$ & 20\\
  BurnbackAFC4 & -18.85 & 0.1 & 11.2 & 40 & $\ket{1/2g} \to \ket{5/2e}$ & 20\\
  BurnbackAFC5 & -21.15 & 0.1 & 11.2 & 40 & $\ket{1/2g} \to \ket{5/2e}$ & 20\\
  BurnbackAFC6 & -23.45 & 0.1 & 11.2 & 40 & $\ket{1/2g} \to \ket{5/2e}$ & 20\\
 \hline
 \hline
\end{tabular}
}
\caption{The $\nu$ column lists the relative center frequencies of the pulses. The frequencies are defined with respect to the $\ket{1/2g} \to \ket{1/2e}$ transition, for an arbitrary ion class, which is set to be at zero. The light intensity for each pulse was chosen such that it matches the Rabi frequency to the relative oscillator strength of the target transitions, shown in the sixth column. We note that the `cleaning' pulses utilized a hybrid pulse combining the edges of a sechyp pulse with a linear frequency chirp. The parameters indicated in the table pertain to the sechyp parameters, whereas the linear frequency scan was 800 kHz for all cleaning pulses. The Reps column shows the number of repetitions used for each of the pulses used.} 
\end{table}
More specifically, the first two pulses, BurnbackAFC1 and BurnbackAFC2, transferred two distinct groups of ions, separated by 2.3 MHz, to the $\ket{3/2g}$ state. Each of the two groups of ions contributed to two peaks of the comb structure, separated by 4.6 MHz, corresponding to the $\ket{3/2g} \to \ket{1/2e}$ and $\ket{3/2g} \to \ket{3/2e}$ transitions. The resulting peaks are shown in Fig.\ \ref{fig:afc_struct}, where the solid (dashed) orange lines indicate ions absorbing to the $\ket{1/2e}$ ($\ket{3/2e}$) state. In addition to the targeted ions, these pulses also burned back multiple undesired $\ket{1/2g}$ ions. These undesired ions were removed using the CleanAFC1, CleanAFC2, CleanAFC3, and CleanAFC4 pulses. In order to further increase the peak absorption of the comb structure, four more groups of ions, with the $\ket{5/2g} \to \ket{5/2e}$ transition frequency coincident with the already prepared $\ket{3/2g}$ peaks, were burned back using the pulses BurnbackAFC3, BurnbackAFC4, BurnbackAFC5, and BurnbackAFC6. The resulting $\ket{5/2g} \to \ket{5/2e}$ transitions are represented by the dotted green line in Fig.\ \ref{fig:afc_struct}. 

\section{AFC peak linewidth and optical depth \label{sec:readout}}

In order to fit the efficiency model (Eq.\ (3) of the main text) to the experimental data, the peak width of the created comb structure has to be determined. Due to the large optical depth of the comb structure, such a characterization is challenging. More specifically, for light polarized parallel to the $D_2$ optical extinction axis the absorption coefficient for Pr$^{3+}$:Y$_2$SiO$_5$ doped at a 500 ppm level has been determined to be $47 \pm 5$ cm$^{-1}$ \cite{sun_rare_2006}. Consequently, even if the peak height is only 20\% of the total optical depth, an absorption measurement designed to not re-pump the atomic ensemble would lead to a total absorption of the probe light. One approach to make this problem more tractable is to prepare the AFC structure using light polarized parallel to the $D_2$ axis followed by a probe beam with the polarization rotated parallel to the $D_1$ axis. Since the absorption coefficient of 500 ppm Pr$^{3+}$:Y$_2$SiO$_5$ for light polarized parallel to the $D_1$ axis is $3.6 \pm 0.5$ cm$^{-1}$, this somewhat reduces the issues inherent with a highly absorbing sample, while still allowing one to infer the peak absorption for light with $D_2$ polarization. Nevertheless, an absorption measurement of such narrow peaks entails considerable additional complexity; in particular, to resolve the comb structure a coherent readout method is utilized. This method consists of a chirped pulse scanned at a rate of 1 MHz/$\mu$s, which places absorbing atoms into a superposition state between the ground and the excited state. The atoms subsequently radiate coherently into the mode of the probe pulse via a free-induction decay, and the emitted photons interfere with the readout light that is being continuously detuned in frequency. This results in a coherent beat pattern that may be deconvolved to recover the shape of the atomic ensemble \cite{chang_2005}.

\begin{figure}[tb!]
\centering
\includegraphics[width=\columnwidth]{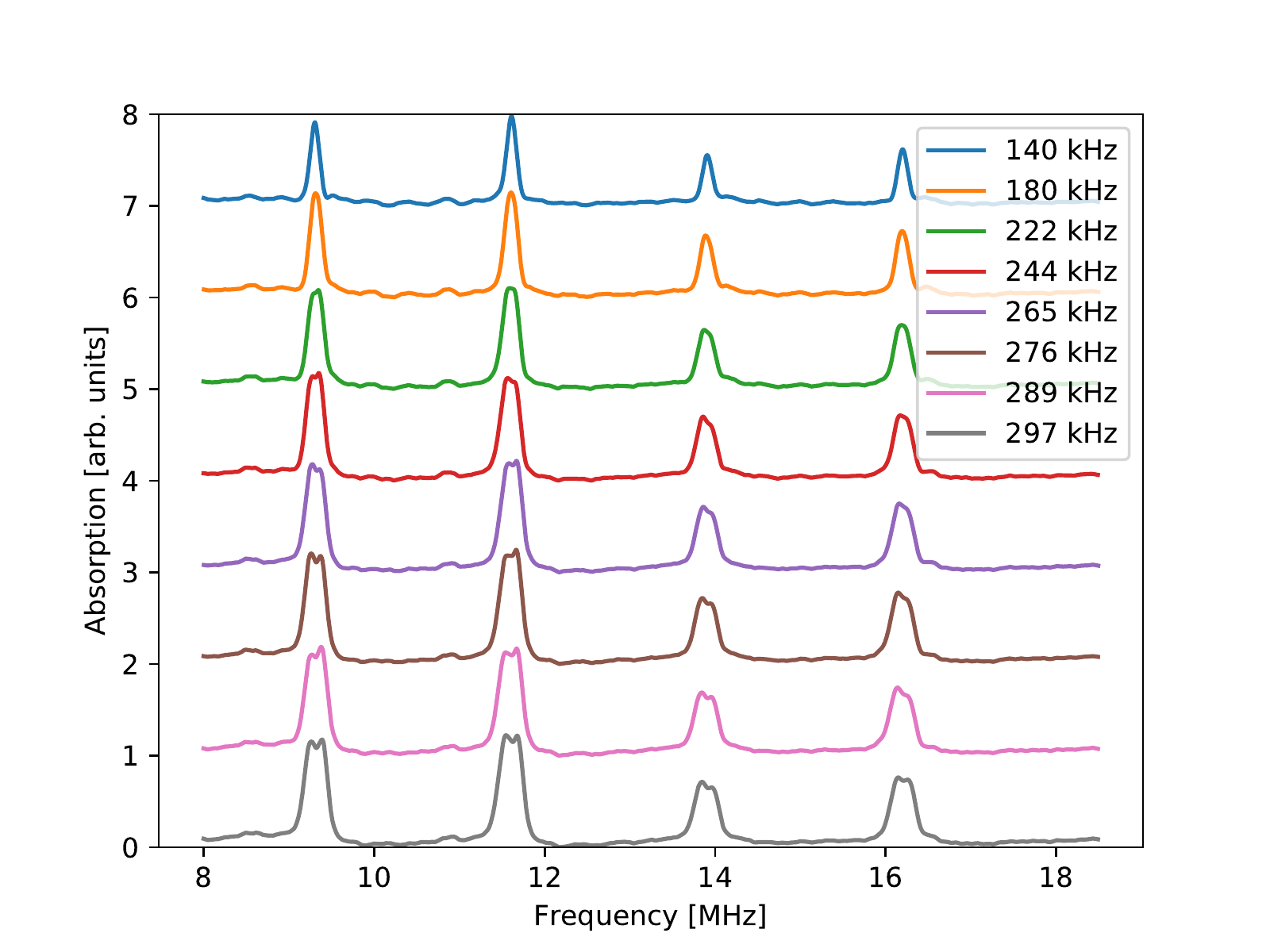}
\caption{Absorption measurement for a range of comb peak linewidth using $D_1$ polarization.}
\label{fig:pw_readout}
\end{figure}

The above outlined absorption measurement was implemented in order to characterize the AFC structure, which required a beam propagation along the crystallographic $b$ axis. However, since the crystal surfaces normal to the $b$ axis contained two pairs of gold-coated electrodes, the comb creation sequence was tested in a different Pr$^{3+}$:Y$_2$SiO$_5$ sample with a nominally equivalent praseodymium concentration, and a length of 12 mm along the $b$ axis. 
\begin{figure}[h]
\centering
\includegraphics[width=\columnwidth]{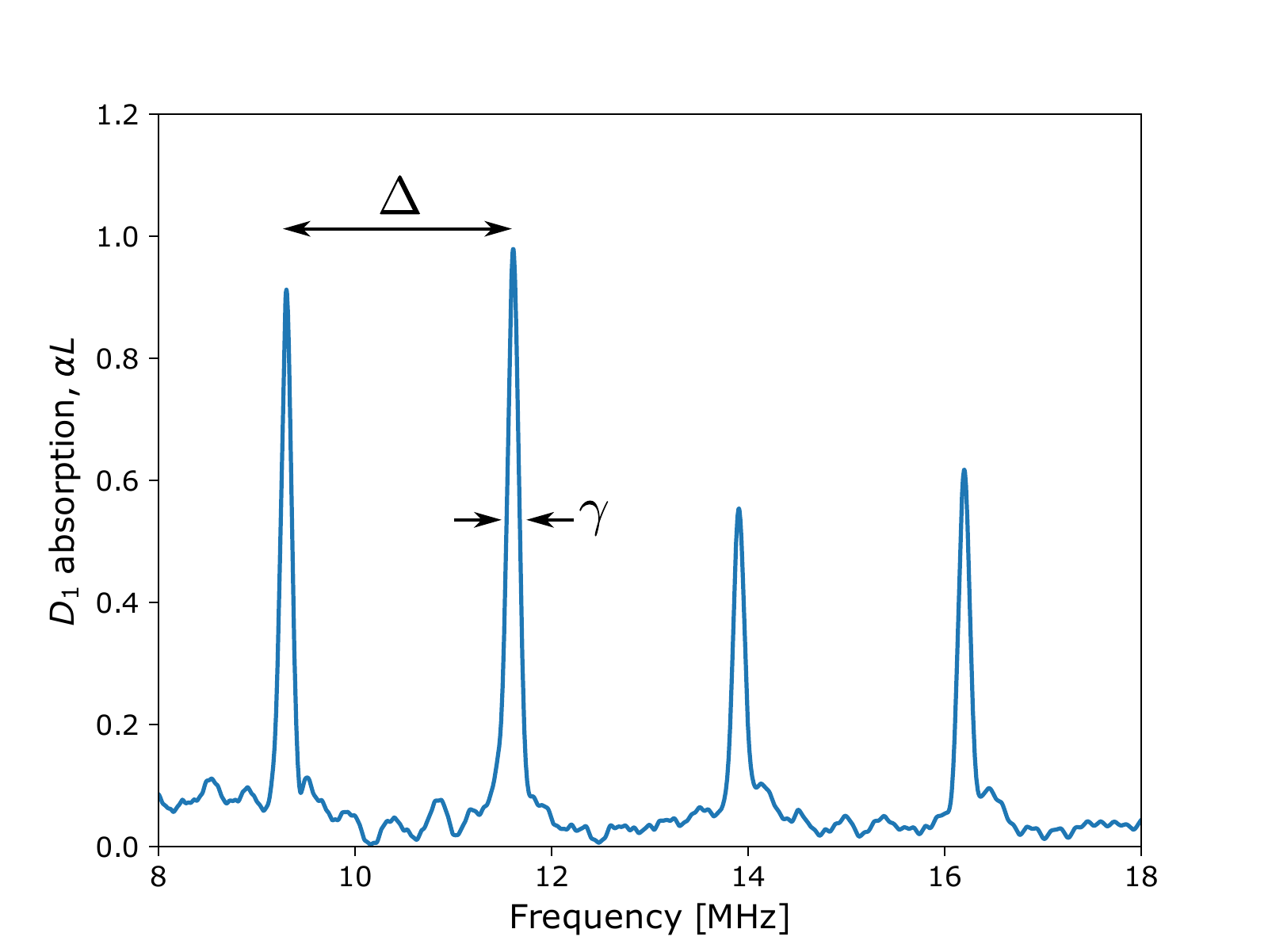}
\caption{An absorption measurement of the AFC structure, performed with the polarization rotated parallel to the $D_1$ axis after the comb preparation was performed using a $D_2$ polarization.}
\label{fig:pw_readout_140}
\end{figure}
Figure \ref{fig:pw_readout_140} shows the absorption structure obtained by the pulse sequence shown in Tab.\ \ref{tab:BurnbackAFC} for a sechyp width $\nu_{\mathrm{width}} = 100$ kHz. By fitting Gaussian lineshapes to the peaks for all four AFC peaks from multiple measurement shots, we estimate the uncertainty of the linewidth to be approximately $\pm 5$ \%. While this assumes Gaussian peak ensembles, this level of accuracy is sufficient for the purpose of corroborating the order-of-magnitude accuracy of the efficiency model. We note that the raw data slightly overestimated the peak width due to a bandwidth limitation of our detection setup.  The light transmitted through the sample was detected using a silicon photodiode in conjunction with a Femto DHPCA-100 trans-impedance amplifier, which, in order to detect the weak probe light, necessitated a gain of $10^5$, which corresponded to a bandwidth of 3.5 MHz. While this limit is not severe and only leads to a small over-estimation (a reduction from 150 kHz to 140 kHz for the narrowest peak), the data in Fig.\ \ref{fig:pw_readout} has been adjusted for this limitation. The appropriate compensation factor was obtained by measuring the complex detector response and simulating the effect on the coherent-readout method. 

Turning to the optical depth, the above measurement yielded a peak absorption of approximately $\alpha L = 0.8$ for $D_1$ polarization in the 12 mm sample. From this, one can infer that the corresponding $D_2$ absorption in the crystal utilized during the memory experiments would be approximately $\alpha L \approx 9$. This is considerably below the absorption of $\alpha L = 45$ cm$^{-1}$ obtained, in the main text, from the recall efficiency model for a peak FWHM of 140 kHz. The most likely explanation for this difference is a shift of the center frequency of the inhomogeneous absorption profile and/or broadening of the absorption profile (due to different sample mounting or inherent sample differences) with respect to the laser locking frequency. The locking point was specifically optimized for maximum absorption in the memory crystal, and the same locking point was utilized for the readout measurements in the 12 mm crystal. There are two additional factors that could also contribute to observed optical depth disagreement. First, the use of a different sample adds uncertainty, both because of the possibility of differences in the doping concentration, as well as growth conditions. Second, the coherent readout method may also underestimate the optical depth, since, even for a peak absorption of  $\alpha L \approx 0.8$, the emitted free-induction decay photons will be reabsorbed to some extent. 

As an independent corroboration of the optical depth obtained from the efficiency model, we note that using only moderate transfer efficiencies of $5\%$ per pulse, the simulations from Sec.\ \ref{sec:afc_burn} yield a peak height comparable to the sides of the 18 MHz transmission window. This has been determined to be $47 \pm 5$ cm$^{-1}$ \cite{sun_rare_2006}, and therefore comes close to the model peak absorption of $45$. From this, in conjunction with the above outlined limitations of the absorption measurement, we will assume that the peak absorption estimated from the model is of the correct magnitude. 

A separate point apparent from Fig.\ \ref{fig:pw_readout_140} is that there is a noticeable difference in the height of the four peaks. In particular, the two low-frequency peaks are higher. We attribute this to the comb creation pulses used for the $\ket{3/2g}$ ions, in conjunction with the cleaning pulses. In particular, the simulation result presented in Fig. \ref{fig:afc_struct} shows that the sequence given in Tab.\ \ref{tab:BurnbackAFC} yields some ions resonant with the $\ket{3/2g} \to \ket{3/2e}$ transition that are overlapping with the two low-frequency comb peaks. The $\ket{3/2g} \to \ket{1/2e}$ transition of these ions lies outside the comb bandwidth, and therefore do not contribute, thus leading to a somewhat higher absorption for the two low-frequency peaks. 

In order to obtain an experimental linewidth for a range of comb-peak linewdiths, the above absorption depth measurement was repeated using several sechyp widths  $\nu_{\mathrm{width}}$. This full set of measurements is shown in Fig. \ref{fig:pw_readout}.

\section{Variable peak-width AFC efficiency} \label{sec:peak width}

\begin{figure}[tbh!]
\centering
\includegraphics[width=\textwidth]{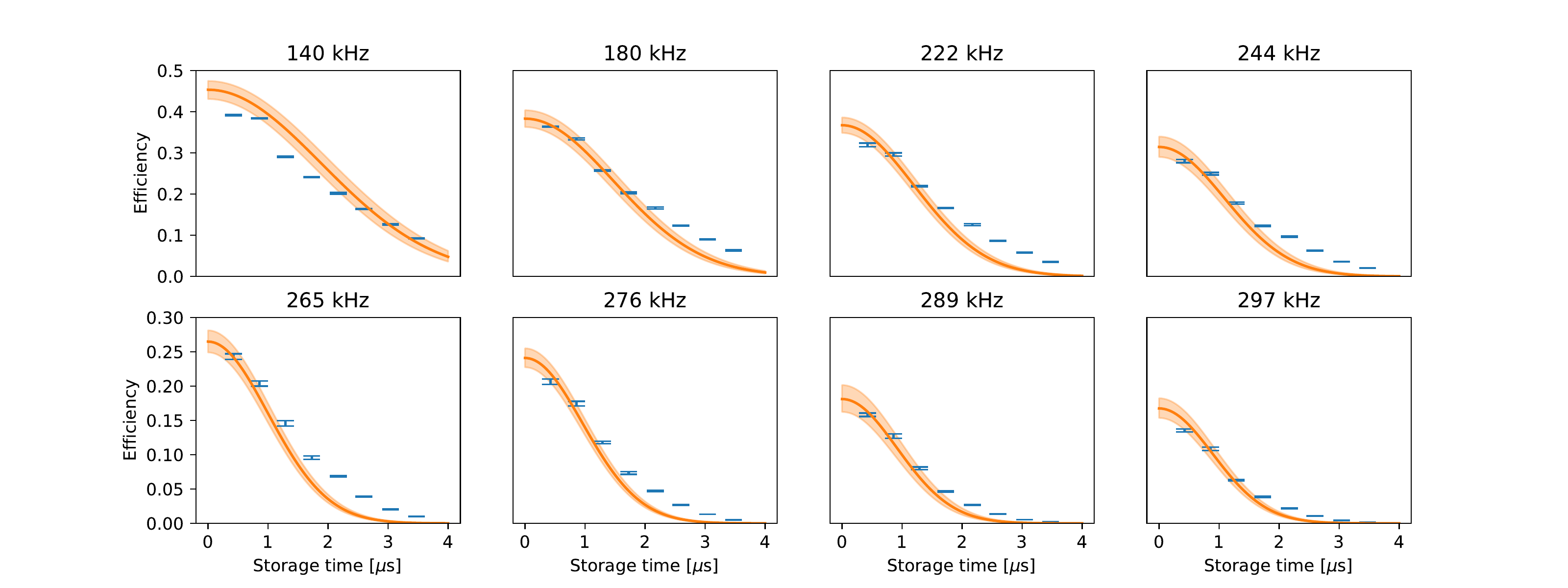}
\caption{The complete set of peak-width measurements overlaid with the efficiency estimated using Eq.\ (3) of the main text. Uncertainties of measured efficiencies correspond to the standard error, while the model efficiency is calculated form a 5\% uncertainty in the peak linewidth.}
\label{fig:eff_plots}
\end{figure}

To modify the peak width, the complex hyperbolic secant pulses employed for peak creation, as described in Sec.\ \ref{sec:afc_burn}, were adjusted to have a larger $\nu_{\mathrm{width}}$, and the intensity scaled to achieve a similar peak $\alpha$. Figure \ref{fig:eff_plots} shows the complete data set for the memory efficiency with respect to a range of AFC peak widths. The efficiency was estimated by comparing the recall intensity of the first echo with a pulse of equal magnitude propagated through the 18 MHz transmission window without any absorbing structure. Such a baseline transmission was previously verified to be within a few percent of the transmission through our optical setup without the crystal. The solid lines correspond to Eq.\ (3) of the main text. The measured efficiency errors correspond to the standard error, while the model error stems from the $\pm 5\%$ uncertainty of the peak width.

Due to the uncertainty of the absorption measurement, as discussed in Sec.\ \ref{sec:readout}, the model $\alpha$ was optimized to achieve a good fit. It was found to range from  $\alpha L = 45$ for the 140 kHz peak width to $\alpha L =  34$ for the 297 kHz peak width. 

\onecolumngrid

\twocolumngrid


\begin{thebibliography}{45}%
\makeatletter
\providecommand \@ifxundefined [1]{%
 \@ifx{#1\undefined}
}%
\providecommand \@ifnum [1]{%
 \ifnum #1\expandafter \@firstoftwo
 \else \expandafter \@secondoftwo
 \fi
}%
\providecommand \@ifx [1]{%
 \ifx #1\expandafter \@firstoftwo
 \else \expandafter \@secondoftwo
 \fi
}%
\providecommand \natexlab [1]{#1}%
\providecommand \enquote  [1]{``#1''}%
\providecommand \bibnamefont  [1]{#1}%
\providecommand \bibfnamefont [1]{#1}%
\providecommand \citenamefont [1]{#1}%
\providecommand \href@noop [0]{\@secondoftwo}%
\providecommand \href [0]{\begingroup \@sanitize@url \@href}%
\providecommand \@href[1]{\@@startlink{#1}\@@href}%
\providecommand \@@href[1]{\endgroup#1\@@endlink}%
\providecommand \@sanitize@url [0]{\catcode `\\12\catcode `\$12\catcode
  `\&12\catcode `\#12\catcode `\^12\catcode `\_12\catcode `\%12\relax}%
\providecommand \@@startlink[1]{}%
\providecommand \@@endlink[0]{}%
\providecommand \url  [0]{\begingroup\@sanitize@url \@url }%
\providecommand \@url [1]{\endgroup\@href {#1}{\urlprefix }}%
\providecommand \urlprefix  [0]{URL }%
\providecommand \Eprint [0]{\href }%
\providecommand \doibase [0]{http://dx.doi.org/}%
\providecommand \selectlanguage [0]{\@gobble}%
\providecommand \bibinfo  [0]{\@secondoftwo}%
\providecommand \bibfield  [0]{\@secondoftwo}%
\providecommand \translation [1]{[#1]}%
\providecommand \BibitemOpen [0]{}%
\providecommand \bibitemStop [0]{}%
\providecommand \bibitemNoStop [0]{.\EOS\space}%
\providecommand \EOS [0]{\spacefactor3000\relax}%
\providecommand \BibitemShut  [1]{\csname bibitem#1\endcsname}%
\let\auto@bib@innerbib\@empty
\bibitem [{\citenamefont {Muralidharan}\ \emph {et~al.}(2016)\citenamefont
  {Muralidharan}, \citenamefont {Li}, \citenamefont {Kim}, \citenamefont
  {Lütkenhaus}, \citenamefont {Lukin},\ and\ \citenamefont
  {Jiang}}]{muralidharan_2016}%
  \BibitemOpen
  \bibfield  {author} {\bibinfo {author} {\bibfnamefont {S.}~\bibnamefont
  {Muralidharan}}, \bibinfo {author} {\bibfnamefont {L.}~\bibnamefont {Li}},
  \bibinfo {author} {\bibfnamefont {J.}~\bibnamefont {Kim}}, \bibinfo {author}
  {\bibfnamefont {N.}~\bibnamefont {Lütkenhaus}}, \bibinfo {author}
  {\bibfnamefont {M.~D.}\ \bibnamefont {Lukin}}, \ and\ \bibinfo {author}
  {\bibfnamefont {L.}~\bibnamefont {Jiang}},\ }\href {\doibase
  10.1038/srep20463} {\bibfield  {journal} {\bibinfo  {journal} {Scientific
  Reports}\ }\textbf {\bibinfo {volume} {6}},\ \bibinfo {pages} {1} (\bibinfo
  {year} {2016})}\BibitemShut {NoStop}%
\bibitem [{\citenamefont {Longdell}\ \emph {et~al.}(2005)\citenamefont
  {Longdell}, \citenamefont {Fraval}, \citenamefont {Sellars},\ and\
  \citenamefont {Manson}}]{longdell_2005}%
  \BibitemOpen
  \bibfield  {author} {\bibinfo {author} {\bibfnamefont {J.~J.}\ \bibnamefont
  {Longdell}}, \bibinfo {author} {\bibfnamefont {E.}~\bibnamefont {Fraval}},
  \bibinfo {author} {\bibfnamefont {M.~J.}\ \bibnamefont {Sellars}}, \ and\
  \bibinfo {author} {\bibfnamefont {N.~B.}\ \bibnamefont {Manson}},\ }\href
  {\doibase 10.1103/PhysRevLett.95.063601} {\bibfield  {journal} {\bibinfo
  {journal} {Physical Review Letters}\ }\textbf {\bibinfo {volume} {95}},\
  \bibinfo {pages} {063601} (\bibinfo {year} {2005})}\BibitemShut {NoStop}%
\bibitem [{\citenamefont {Hsu}\ \emph {et~al.}(2006)\citenamefont {Hsu},
  \citenamefont {Hétet}, \citenamefont {Glöckl}, \citenamefont {Longdell},
  \citenamefont {Buchler}, \citenamefont {Bachor},\ and\ \citenamefont
  {Lam}}]{hsu_2006}%
  \BibitemOpen
  \bibfield  {author} {\bibinfo {author} {\bibfnamefont {M.~T.~L.}\
  \bibnamefont {Hsu}}, \bibinfo {author} {\bibfnamefont {G.}~\bibnamefont
  {Hétet}}, \bibinfo {author} {\bibfnamefont {O.}~\bibnamefont {Glöckl}},
  \bibinfo {author} {\bibfnamefont {J.~J.}\ \bibnamefont {Longdell}}, \bibinfo
  {author} {\bibfnamefont {B.~C.}\ \bibnamefont {Buchler}}, \bibinfo {author}
  {\bibfnamefont {H.-A.}\ \bibnamefont {Bachor}}, \ and\ \bibinfo {author}
  {\bibfnamefont {P.~K.}\ \bibnamefont {Lam}},\ }\href {\doibase
  10.1103/PhysRevLett.97.183601} {\bibfield  {journal} {\bibinfo  {journal}
  {Physical Review Letters}\ }\textbf {\bibinfo {volume} {97}},\ \bibinfo
  {pages} {183601} (\bibinfo {year} {2006})}\BibitemShut {NoStop}%
\bibitem [{\citenamefont {Schraft}\ \emph {et~al.}(2016)\citenamefont
  {Schraft}, \citenamefont {Hain}, \citenamefont {Lorenz},\ and\ \citenamefont
  {Halfmann}}]{schraft_2016}%
  \BibitemOpen
  \bibfield  {author} {\bibinfo {author} {\bibfnamefont {D.}~\bibnamefont
  {Schraft}}, \bibinfo {author} {\bibfnamefont {M.}~\bibnamefont {Hain}},
  \bibinfo {author} {\bibfnamefont {N.}~\bibnamefont {Lorenz}}, \ and\ \bibinfo
  {author} {\bibfnamefont {T.}~\bibnamefont {Halfmann}},\ }\href {\doibase
  10.1103/PhysRevLett.116.073602} {\bibfield  {journal} {\bibinfo  {journal}
  {Physical Review Letters}\ }\textbf {\bibinfo {volume} {116}},\ \bibinfo
  {pages} {073602} (\bibinfo {year} {2016})}\BibitemShut {NoStop}%
\bibitem [{\citenamefont {Reim}\ \emph {et~al.}(2011)\citenamefont {Reim},
  \citenamefont {Michelberger}, \citenamefont {Lee}, \citenamefont {Nunn},
  \citenamefont {Langford},\ and\ \citenamefont {Walmsley}}]{reim_2011}%
  \BibitemOpen
  \bibfield  {author} {\bibinfo {author} {\bibfnamefont {K.~F.}\ \bibnamefont
  {Reim}}, \bibinfo {author} {\bibfnamefont {P.}~\bibnamefont {Michelberger}},
  \bibinfo {author} {\bibfnamefont {K.~C.}\ \bibnamefont {Lee}}, \bibinfo
  {author} {\bibfnamefont {J.}~\bibnamefont {Nunn}}, \bibinfo {author}
  {\bibfnamefont {N.~K.}\ \bibnamefont {Langford}}, \ and\ \bibinfo {author}
  {\bibfnamefont {I.~A.}\ \bibnamefont {Walmsley}},\ }\href {\doibase
  10.1103/PhysRevLett.107.053603} {\bibfield  {journal} {\bibinfo  {journal}
  {Physical Review Letters}\ }\textbf {\bibinfo {volume} {107}},\ \bibinfo
  {pages} {053603} (\bibinfo {year} {2011})}\BibitemShut {NoStop}%
\bibitem [{\citenamefont {Sprague}\ \emph {et~al.}(2014)\citenamefont
  {Sprague}, \citenamefont {Michelberger}, \citenamefont {Champion},
  \citenamefont {England}, \citenamefont {Nunn}, \citenamefont {Jin},
  \citenamefont {Kolthammer}, \citenamefont {Abdolvand}, \citenamefont
  {Russell},\ and\ \citenamefont {Walmsley}}]{sprague_2014}%
  \BibitemOpen
  \bibfield  {author} {\bibinfo {author} {\bibfnamefont {M.~R.}\ \bibnamefont
  {Sprague}}, \bibinfo {author} {\bibfnamefont {P.~S.}\ \bibnamefont
  {Michelberger}}, \bibinfo {author} {\bibfnamefont {T.~F.~M.}\ \bibnamefont
  {Champion}}, \bibinfo {author} {\bibfnamefont {D.~G.}\ \bibnamefont
  {England}}, \bibinfo {author} {\bibfnamefont {J.}~\bibnamefont {Nunn}},
  \bibinfo {author} {\bibfnamefont {X.-M.}\ \bibnamefont {Jin}}, \bibinfo
  {author} {\bibfnamefont {W.~S.}\ \bibnamefont {Kolthammer}}, \bibinfo
  {author} {\bibfnamefont {A.}~\bibnamefont {Abdolvand}}, \bibinfo {author}
  {\bibfnamefont {P.~S.~J.}\ \bibnamefont {Russell}}, \ and\ \bibinfo {author}
  {\bibfnamefont {I.~A.}\ \bibnamefont {Walmsley}},\ }\href {\doibase
  10.1038/nphoton.2014.45} {\bibfield  {journal} {\bibinfo  {journal} {Nature
  Photonics}\ }\textbf {\bibinfo {volume} {8}},\ \bibinfo {pages} {287}
  (\bibinfo {year} {2014})}\BibitemShut {NoStop}%
\bibitem [{\citenamefont {Nilsson}\ and\ \citenamefont
  {Kröll}(2005)}]{nilsson_2005}%
  \BibitemOpen
  \bibfield  {author} {\bibinfo {author} {\bibfnamefont {M.}~\bibnamefont
  {Nilsson}}\ and\ \bibinfo {author} {\bibfnamefont {S.}~\bibnamefont
  {Kröll}},\ }\href {\doibase 10.1016/j.optcom.2004.11.077} {\bibfield
  {journal} {\bibinfo  {journal} {Optics Communications}\ }\textbf {\bibinfo
  {volume} {247}},\ \bibinfo {pages} {393} (\bibinfo {year}
  {2005})}\BibitemShut {NoStop}%
\bibitem [{\citenamefont {Hétet}\ \emph {et~al.}(2008)\citenamefont {Hétet},
  \citenamefont {Longdell}, \citenamefont {Sellars}, \citenamefont {Lam},\ and\
  \citenamefont {Buchler}}]{hetet_2008}%
  \BibitemOpen
  \bibfield  {author} {\bibinfo {author} {\bibfnamefont {G.}~\bibnamefont
  {Hétet}}, \bibinfo {author} {\bibfnamefont {J.~J.}\ \bibnamefont
  {Longdell}}, \bibinfo {author} {\bibfnamefont {M.~J.}\ \bibnamefont
  {Sellars}}, \bibinfo {author} {\bibfnamefont {P.~K.}\ \bibnamefont {Lam}}, \
  and\ \bibinfo {author} {\bibfnamefont {B.~C.}\ \bibnamefont {Buchler}},\
  }\href {\doibase 10.1103/PhysRevLett.101.203601} {\bibfield  {journal}
  {\bibinfo  {journal} {Physical Review Letters}\ }\textbf {\bibinfo {volume}
  {101}},\ \bibinfo {pages} {203601} (\bibinfo {year} {2008})}\BibitemShut
  {NoStop}%
\bibitem [{\citenamefont {Hosseini}\ \emph {et~al.}(2011)\citenamefont
  {Hosseini}, \citenamefont {Sparkes}, \citenamefont {Campbell}, \citenamefont
  {Lam},\ and\ \citenamefont {Buchler}}]{hosseini_2011}%
  \BibitemOpen
  \bibfield  {author} {\bibinfo {author} {\bibfnamefont {M.}~\bibnamefont
  {Hosseini}}, \bibinfo {author} {\bibfnamefont {B.~M.}\ \bibnamefont
  {Sparkes}}, \bibinfo {author} {\bibfnamefont {G.}~\bibnamefont {Campbell}},
  \bibinfo {author} {\bibfnamefont {P.~K.}\ \bibnamefont {Lam}}, \ and\
  \bibinfo {author} {\bibfnamefont {B.~C.}\ \bibnamefont {Buchler}},\ }\href
  {\doibase 10.1038/ncomms1175} {\bibfield  {journal} {\bibinfo  {journal}
  {Nature Communications}\ }\textbf {\bibinfo {volume} {2}},\ \bibinfo {pages}
  {1} (\bibinfo {year} {2011})}\BibitemShut {NoStop}%
\bibitem [{\citenamefont {Hedges}\ \emph {et~al.}(2010)\citenamefont {Hedges},
  \citenamefont {Longdell}, \citenamefont {Li},\ and\ \citenamefont
  {Sellars}}]{hedges_2010}%
  \BibitemOpen
  \bibfield  {author} {\bibinfo {author} {\bibfnamefont {M.~P.}\ \bibnamefont
  {Hedges}}, \bibinfo {author} {\bibfnamefont {J.~J.}\ \bibnamefont
  {Longdell}}, \bibinfo {author} {\bibfnamefont {Y.}~\bibnamefont {Li}}, \ and\
  \bibinfo {author} {\bibfnamefont {M.~J.}\ \bibnamefont {Sellars}},\ }\href
  {\doibase 10.1038/nature09081} {\bibfield  {journal} {\bibinfo  {journal}
  {Nature}\ }\textbf {\bibinfo {volume} {465}},\ \bibinfo {pages} {1052}
  (\bibinfo {year} {2010})}\BibitemShut {NoStop}%
\bibitem [{\citenamefont {Cho}\ \emph {et~al.}(2016)\citenamefont {Cho},
  \citenamefont {Campbell}, \citenamefont {Everett}, \citenamefont {Bernu},
  \citenamefont {Higginbottom}, \citenamefont {Cao}, \citenamefont {Geng},
  \citenamefont {Robins}, \citenamefont {Lam},\ and\ \citenamefont
  {Buchler}}]{cho_2016}%
  \BibitemOpen
  \bibfield  {author} {\bibinfo {author} {\bibfnamefont {Y.-W.}\ \bibnamefont
  {Cho}}, \bibinfo {author} {\bibfnamefont {G.~T.}\ \bibnamefont {Campbell}},
  \bibinfo {author} {\bibfnamefont {J.~L.}\ \bibnamefont {Everett}}, \bibinfo
  {author} {\bibfnamefont {J.}~\bibnamefont {Bernu}}, \bibinfo {author}
  {\bibfnamefont {D.~B.}\ \bibnamefont {Higginbottom}}, \bibinfo {author}
  {\bibfnamefont {M.~T.}\ \bibnamefont {Cao}}, \bibinfo {author} {\bibfnamefont
  {J.}~\bibnamefont {Geng}}, \bibinfo {author} {\bibfnamefont {N.~P.}\
  \bibnamefont {Robins}}, \bibinfo {author} {\bibfnamefont {P.~K.}\
  \bibnamefont {Lam}}, \ and\ \bibinfo {author} {\bibfnamefont {B.~C.}\
  \bibnamefont {Buchler}},\ }\href {\doibase 10.1364/OPTICA.3.000100}
  {\bibfield  {journal} {\bibinfo  {journal} {Optica}\ }\textbf {\bibinfo
  {volume} {3}},\ \bibinfo {pages} {100} (\bibinfo {year} {2016})}\BibitemShut
  {NoStop}%
\bibitem [{\citenamefont {Dajczgewand}\ \emph {et~al.}(2014)\citenamefont
  {Dajczgewand}, \citenamefont {Gou\"{e}t}, \citenamefont {Louchet-Chauvet},\
  and\ \citenamefont {Chaneli\`{e}re}}]{Dajczgewand_ROSE_2014}%
  \BibitemOpen
  \bibfield  {author} {\bibinfo {author} {\bibfnamefont {J.}~\bibnamefont
  {Dajczgewand}}, \bibinfo {author} {\bibfnamefont {J.-L.~L.}\ \bibnamefont
  {Gou\"{e}t}}, \bibinfo {author} {\bibfnamefont {A.}~\bibnamefont
  {Louchet-Chauvet}}, \ and\ \bibinfo {author} {\bibfnamefont {T.}~\bibnamefont
  {Chaneli\`{e}re}},\ }\href {\doibase 10.1364/OL.39.002711} {\bibfield
  {journal} {\bibinfo  {journal} {Opt. Lett.}\ }\textbf {\bibinfo {volume}
  {39}},\ \bibinfo {pages} {2711} (\bibinfo {year} {2014})}\BibitemShut
  {NoStop}%
\bibitem [{\citenamefont {Kutluer}\ \emph {et~al.}(2016)\citenamefont
  {Kutluer}, \citenamefont {Pascual-Winter}, \citenamefont {Dajczgewand},
  \citenamefont {Ledingham}, \citenamefont {Mazzera}, \citenamefont
  {Chaneli\`ere},\ and\ \citenamefont {de~Riedmatten}}]{Kutluer_Holes_2016}%
  \BibitemOpen
  \bibfield  {author} {\bibinfo {author} {\bibfnamefont {K.}~\bibnamefont
  {Kutluer}}, \bibinfo {author} {\bibfnamefont {M.~F.}\ \bibnamefont
  {Pascual-Winter}}, \bibinfo {author} {\bibfnamefont {J.}~\bibnamefont
  {Dajczgewand}}, \bibinfo {author} {\bibfnamefont {P.~M.}\ \bibnamefont
  {Ledingham}}, \bibinfo {author} {\bibfnamefont {M.}~\bibnamefont {Mazzera}},
  \bibinfo {author} {\bibfnamefont {T.}~\bibnamefont {Chaneli\`ere}}, \ and\
  \bibinfo {author} {\bibfnamefont {H.}~\bibnamefont {de~Riedmatten}},\ }\href
  {\doibase 10.1103/PhysRevA.93.040302} {\bibfield  {journal} {\bibinfo
  {journal} {Phys. Rev. A}\ }\textbf {\bibinfo {volume} {93}},\ \bibinfo
  {pages} {040302} (\bibinfo {year} {2016})}\BibitemShut {NoStop}%
\bibitem [{\citenamefont {McAuslan}\ \emph {et~al.}(2011)\citenamefont
  {McAuslan}, \citenamefont {Ledingham}, \citenamefont {Naylor}, \citenamefont
  {Beavan}, \citenamefont {Hedges}, \citenamefont {Sellars},\ and\
  \citenamefont {Longdell}}]{McAuslan2011_HYPER}%
  \BibitemOpen
  \bibfield  {author} {\bibinfo {author} {\bibfnamefont {D.~L.}\ \bibnamefont
  {McAuslan}}, \bibinfo {author} {\bibfnamefont {P.~M.}\ \bibnamefont
  {Ledingham}}, \bibinfo {author} {\bibfnamefont {W.~R.}\ \bibnamefont
  {Naylor}}, \bibinfo {author} {\bibfnamefont {S.~E.}\ \bibnamefont {Beavan}},
  \bibinfo {author} {\bibfnamefont {M.~P.}\ \bibnamefont {Hedges}}, \bibinfo
  {author} {\bibfnamefont {M.~J.}\ \bibnamefont {Sellars}}, \ and\ \bibinfo
  {author} {\bibfnamefont {J.~J.}\ \bibnamefont {Longdell}},\ }\href {\doibase
  10.1103/PhysRevA.84.022309} {\bibfield  {journal} {\bibinfo  {journal} {Phys.
  Rev. A}\ }\textbf {\bibinfo {volume} {84}},\ \bibinfo {pages} {022309}
  (\bibinfo {year} {2011})}\BibitemShut {NoStop}%
\bibitem [{\citenamefont {Afzelius}\ \emph {et~al.}(2009)\citenamefont
  {Afzelius}, \citenamefont {Simon}, \citenamefont {de~Riedmatten},\ and\
  \citenamefont {Gisin}}]{afzelius_multimode_2009}%
  \BibitemOpen
  \bibfield  {author} {\bibinfo {author} {\bibfnamefont {M.}~\bibnamefont
  {Afzelius}}, \bibinfo {author} {\bibfnamefont {C.}~\bibnamefont {Simon}},
  \bibinfo {author} {\bibfnamefont {H.}~\bibnamefont {de~Riedmatten}}, \ and\
  \bibinfo {author} {\bibfnamefont {N.}~\bibnamefont {Gisin}},\ }\href
  {\doibase 10.1103/PhysRevA.79.052329} {\bibfield  {journal} {\bibinfo
  {journal} {Physical Review A}\ }\textbf {\bibinfo {volume} {79}},\ \bibinfo
  {pages} {052329} (\bibinfo {year} {2009})}\BibitemShut {NoStop}%
\bibitem [{\citenamefont {Minář}\ \emph {et~al.}(2009)\citenamefont
  {Minář}, \citenamefont {Lauritzen}, \citenamefont {Riedmatten},
  \citenamefont {Afzelius}, \citenamefont {Simon},\ and\ \citenamefont
  {Gisin}}]{minar_2009}%
  \BibitemOpen
  \bibfield  {author} {\bibinfo {author} {\bibfnamefont {J.}~\bibnamefont
  {Minář}}, \bibinfo {author} {\bibfnamefont {B.}~\bibnamefont {Lauritzen}},
  \bibinfo {author} {\bibfnamefont {H.~d.}\ \bibnamefont {Riedmatten}},
  \bibinfo {author} {\bibfnamefont {M.}~\bibnamefont {Afzelius}}, \bibinfo
  {author} {\bibfnamefont {C.}~\bibnamefont {Simon}}, \ and\ \bibinfo {author}
  {\bibfnamefont {N.}~\bibnamefont {Gisin}},\ }\href {\doibase
  10.1088/1367-2630/11/11/113019} {\bibfield  {journal} {\bibinfo  {journal}
  {New Journal of Physics}\ }\textbf {\bibinfo {volume} {11}},\ \bibinfo
  {pages} {113019} (\bibinfo {year} {2009})}\BibitemShut {NoStop}%
\bibitem [{\citenamefont {Afzelius}\ and\ \citenamefont
  {Simon}(2010)}]{afzelius_impedance_2010}%
  \BibitemOpen
  \bibfield  {author} {\bibinfo {author} {\bibfnamefont {M.}~\bibnamefont
  {Afzelius}}\ and\ \bibinfo {author} {\bibfnamefont {C.}~\bibnamefont
  {Simon}},\ }\href {\doibase 10.1103/PhysRevA.82.022310} {\bibfield  {journal}
  {\bibinfo  {journal} {Physical Review A}\ }\textbf {\bibinfo {volume} {82}},\
  \bibinfo {pages} {022310} (\bibinfo {year} {2010})}\BibitemShut {NoStop}%
\bibitem [{\citenamefont {Sabooni}\ \emph {et~al.}(2013)\citenamefont
  {Sabooni}, \citenamefont {Li}, \citenamefont {Kröll},\ and\ \citenamefont
  {Rippe}}]{sabooni_2013}%
  \BibitemOpen
  \bibfield  {author} {\bibinfo {author} {\bibfnamefont {M.}~\bibnamefont
  {Sabooni}}, \bibinfo {author} {\bibfnamefont {Q.}~\bibnamefont {Li}},
  \bibinfo {author} {\bibfnamefont {S.}~\bibnamefont {Kröll}}, \ and\ \bibinfo
  {author} {\bibfnamefont {L.}~\bibnamefont {Rippe}},\ }\href {\doibase
  10.1103/PhysRevLett.110.133604} {\bibfield  {journal} {\bibinfo  {journal}
  {Physical Review Letters}\ }\textbf {\bibinfo {volume} {110}},\ \bibinfo
  {pages} {133604} (\bibinfo {year} {2013})}\BibitemShut {NoStop}%
\bibitem [{\citenamefont {Jobez}\ \emph {et~al.}(2015)\citenamefont {Jobez},
  \citenamefont {Laplane}, \citenamefont {Timoney}, \citenamefont {Gisin},
  \citenamefont {Ferrier}, \citenamefont {Goldner},\ and\ \citenamefont
  {Afzelius}}]{jobez_2015}%
  \BibitemOpen
  \bibfield  {author} {\bibinfo {author} {\bibfnamefont {P.}~\bibnamefont
  {Jobez}}, \bibinfo {author} {\bibfnamefont {C.}~\bibnamefont {Laplane}},
  \bibinfo {author} {\bibfnamefont {N.}~\bibnamefont {Timoney}}, \bibinfo
  {author} {\bibfnamefont {N.}~\bibnamefont {Gisin}}, \bibinfo {author}
  {\bibfnamefont {A.}~\bibnamefont {Ferrier}}, \bibinfo {author} {\bibfnamefont
  {P.}~\bibnamefont {Goldner}}, \ and\ \bibinfo {author} {\bibfnamefont
  {M.}~\bibnamefont {Afzelius}},\ }\href {\doibase
  10.1103/PhysRevLett.114.230502} {\bibfield  {journal} {\bibinfo  {journal}
  {Physical Review Letters}\ }\textbf {\bibinfo {volume} {114}},\ \bibinfo
  {pages} {230502} (\bibinfo {year} {2015})}\BibitemShut {NoStop}%
\bibitem [{\citenamefont {G{\"u}ndo{\u g}an}\ \emph {et~al.}(2015)\citenamefont
  {G{\"u}ndo{\u g}an}, \citenamefont {Ledingham}, \citenamefont {Kutluer},
  \citenamefont {Mazzera},\ and\ \citenamefont {{de
  Riedmatten}}}]{gundogan2015}%
  \BibitemOpen
  \bibfield  {author} {\bibinfo {author} {\bibfnamefont {M.}~\bibnamefont
  {G{\"u}ndo{\u g}an}}, \bibinfo {author} {\bibfnamefont {P.~M.}\ \bibnamefont
  {Ledingham}}, \bibinfo {author} {\bibfnamefont {K.}~\bibnamefont {Kutluer}},
  \bibinfo {author} {\bibfnamefont {M.}~\bibnamefont {Mazzera}}, \ and\
  \bibinfo {author} {\bibfnamefont {H.}~\bibnamefont {{de Riedmatten}}},\
  }\href {\doibase 10.1103/PhysRevLett.114.230501} {\bibfield  {journal}
  {\bibinfo  {journal} {Physical Review Letters}\ }\textbf {\bibinfo {volume}
  {114}},\ \bibinfo {pages} {230501} (\bibinfo {year} {2015})}\BibitemShut
  {NoStop}%
\bibitem [{\citenamefont {Timoney}\ \emph {et~al.}(2013)\citenamefont
  {Timoney}, \citenamefont {Usmani}, \citenamefont {Jobez}, \citenamefont
  {Afzelius},\ and\ \citenamefont {Gisin}}]{timoney_single_2013}%
  \BibitemOpen
  \bibfield  {author} {\bibinfo {author} {\bibfnamefont {N.}~\bibnamefont
  {Timoney}}, \bibinfo {author} {\bibfnamefont {I.}~\bibnamefont {Usmani}},
  \bibinfo {author} {\bibfnamefont {P.}~\bibnamefont {Jobez}}, \bibinfo
  {author} {\bibfnamefont {M.}~\bibnamefont {Afzelius}}, \ and\ \bibinfo
  {author} {\bibfnamefont {N.}~\bibnamefont {Gisin}},\ }\href {\doibase
  10.1103/PhysRevA.88.022324} {\bibfield  {journal} {\bibinfo  {journal}
  {Physical Review A}\ }\textbf {\bibinfo {volume} {88}},\ \bibinfo {pages}
  {022324} (\bibinfo {year} {2013})}\BibitemShut {NoStop}%
\bibitem [{\citenamefont {Bonarota}\ \emph {et~al.}(2014)\citenamefont
  {Bonarota}, \citenamefont {Dajczgewand}, \citenamefont {Louchet-Chauvet},
  \citenamefont {Gouët},\ and\ \citenamefont {Chanelière}}]{bonarota_2014}%
  \BibitemOpen
  \bibfield  {author} {\bibinfo {author} {\bibfnamefont {M.}~\bibnamefont
  {Bonarota}}, \bibinfo {author} {\bibfnamefont {J.}~\bibnamefont
  {Dajczgewand}}, \bibinfo {author} {\bibfnamefont {A.}~\bibnamefont
  {Louchet-Chauvet}}, \bibinfo {author} {\bibfnamefont {J.-L.~L.}\ \bibnamefont
  {Gouët}}, \ and\ \bibinfo {author} {\bibfnamefont {T.}~\bibnamefont
  {Chanelière}},\ }\href {\doibase 10.1088/1054-660X/24/9/094003} {\bibfield
  {journal} {\bibinfo  {journal} {Laser Physics}\ }\textbf {\bibinfo {volume}
  {24}},\ \bibinfo {pages} {094003} (\bibinfo {year} {2014})}\BibitemShut
  {NoStop}%
\bibitem [{\citenamefont {Arcangeli}\ \emph {et~al.}(2016)\citenamefont
  {Arcangeli}, \citenamefont {Ferrier},\ and\ \citenamefont
  {Goldner}}]{arcangeli_2016}%
  \BibitemOpen
  \bibfield  {author} {\bibinfo {author} {\bibfnamefont {A.}~\bibnamefont
  {Arcangeli}}, \bibinfo {author} {\bibfnamefont {A.}~\bibnamefont {Ferrier}},
  \ and\ \bibinfo {author} {\bibfnamefont {P.}~\bibnamefont {Goldner}},\ }\href
  {\doibase 10.1103/PhysRevA.93.062303} {\bibfield  {journal} {\bibinfo
  {journal} {Physical Review A}\ }\textbf {\bibinfo {volume} {93}},\ \bibinfo
  {pages} {062303} (\bibinfo {year} {2016})}\BibitemShut {NoStop}%
\bibitem [{\citenamefont {Lobino}\ \emph {et~al.}(2009)\citenamefont {Lobino},
  \citenamefont {Kupchak}, \citenamefont {Figueroa},\ and\ \citenamefont
  {Lvovsky}}]{lobino_2009}%
  \BibitemOpen
  \bibfield  {author} {\bibinfo {author} {\bibfnamefont {M.}~\bibnamefont
  {Lobino}}, \bibinfo {author} {\bibfnamefont {C.}~\bibnamefont {Kupchak}},
  \bibinfo {author} {\bibfnamefont {E.}~\bibnamefont {Figueroa}}, \ and\
  \bibinfo {author} {\bibfnamefont {A.~I.}\ \bibnamefont {Lvovsky}},\ }\href
  {\doibase 10.1103/PhysRevLett.102.203601} {\bibfield  {journal} {\bibinfo
  {journal} {Physical Review Letters}\ }\textbf {\bibinfo {volume} {102}},\
  \bibinfo {pages} {203601} (\bibinfo {year} {2009})}\BibitemShut {NoStop}%
\bibitem [{\citenamefont {Equall}\ \emph {et~al.}(1994)\citenamefont {Equall},
  \citenamefont {Sun}, \citenamefont {Cone},\ and\ \citenamefont
  {Macfarlane}}]{equall_ultraslow_1994}%
  \BibitemOpen
  \bibfield  {author} {\bibinfo {author} {\bibfnamefont {R.~W.}\ \bibnamefont
  {Equall}}, \bibinfo {author} {\bibfnamefont {Y.}~\bibnamefont {Sun}},
  \bibinfo {author} {\bibfnamefont {R.~L.}\ \bibnamefont {Cone}}, \ and\
  \bibinfo {author} {\bibfnamefont {R.~M.}\ \bibnamefont {Macfarlane}},\ }\href
  {\doibase 10.1103/PhysRevLett.72.2179} {\bibfield  {journal} {\bibinfo
  {journal} {Physical Review Letters}\ }\textbf {\bibinfo {volume} {72}},\
  \bibinfo {pages} {2179} (\bibinfo {year} {1994})}\BibitemShut {NoStop}%
\bibitem [{\citenamefont {Graf}\ \emph {et~al.}(1997)\citenamefont {Graf},
  \citenamefont {Renn}, \citenamefont {Wild},\ and\ \citenamefont
  {Mitsunaga}}]{graf_1997}%
  \BibitemOpen
  \bibfield  {author} {\bibinfo {author} {\bibfnamefont {F.~R.}\ \bibnamefont
  {Graf}}, \bibinfo {author} {\bibfnamefont {A.}~\bibnamefont {Renn}}, \bibinfo
  {author} {\bibfnamefont {U.~P.}\ \bibnamefont {Wild}}, \ and\ \bibinfo
  {author} {\bibfnamefont {M.}~\bibnamefont {Mitsunaga}},\ }\href {\doibase
  10.1103/PhysRevB.55.11225} {\bibfield  {journal} {\bibinfo  {journal}
  {Physical Review B}\ }\textbf {\bibinfo {volume} {55}},\ \bibinfo {pages}
  {11225} (\bibinfo {year} {1997})}\BibitemShut {NoStop}%
\bibitem [{\citenamefont {Beavan}\ \emph {et~al.}(2013)\citenamefont {Beavan},
  \citenamefont {Goldschmidt},\ and\ \citenamefont {Sellars}}]{beavan_2013}%
  \BibitemOpen
  \bibfield  {author} {\bibinfo {author} {\bibfnamefont {S.~E.}\ \bibnamefont
  {Beavan}}, \bibinfo {author} {\bibfnamefont {E.~A.}\ \bibnamefont
  {Goldschmidt}}, \ and\ \bibinfo {author} {\bibfnamefont {M.~J.}\ \bibnamefont
  {Sellars}},\ }\href {\doibase 10.1364/JOSAB.30.001173} {\bibfield  {journal}
  {\bibinfo  {journal} {JOSA B}\ }\textbf {\bibinfo {volume} {30}},\ \bibinfo
  {pages} {1173} (\bibinfo {year} {2013})}\BibitemShut {NoStop}%
\bibitem [{\citenamefont {Lauritzen}\ \emph {et~al.}(2011)\citenamefont
  {Lauritzen}, \citenamefont {Minář}, \citenamefont {de~Riedmatten},
  \citenamefont {Afzelius},\ and\ \citenamefont {Gisin}}]{lauritzen_2011}%
  \BibitemOpen
  \bibfield  {author} {\bibinfo {author} {\bibfnamefont {B.}~\bibnamefont
  {Lauritzen}}, \bibinfo {author} {\bibfnamefont {J.}~\bibnamefont {Minář}},
  \bibinfo {author} {\bibfnamefont {H.}~\bibnamefont {de~Riedmatten}}, \bibinfo
  {author} {\bibfnamefont {M.}~\bibnamefont {Afzelius}}, \ and\ \bibinfo
  {author} {\bibfnamefont {N.}~\bibnamefont {Gisin}},\ }\href {\doibase
  10.1103/PhysRevA.83.012318} {\bibfield  {journal} {\bibinfo  {journal}
  {Physical Review A}\ }\textbf {\bibinfo {volume} {83}},\ \bibinfo {pages}
  {012318} (\bibinfo {year} {2011})}\BibitemShut {NoStop}%
\bibitem [{\citenamefont {Wang}\ and\ \citenamefont
  {Meltzer}(1992)}]{wang_1992}%
  \BibitemOpen
  \bibfield  {author} {\bibinfo {author} {\bibfnamefont {Y.~P.}\ \bibnamefont
  {Wang}}\ and\ \bibinfo {author} {\bibfnamefont {R.~S.}\ \bibnamefont
  {Meltzer}},\ }\href {\doibase 10.1103/PhysRevB.45.10119} {\bibfield
  {journal} {\bibinfo  {journal} {Physical Review B}\ }\textbf {\bibinfo
  {volume} {45}},\ \bibinfo {pages} {10119} (\bibinfo {year}
  {1992})}\BibitemShut {NoStop}%
\bibitem [{\citenamefont {Meixner}\ \emph {et~al.}(1992)\citenamefont
  {Meixner}, \citenamefont {Jefferson},\ and\ \citenamefont
  {Macfarlane}}]{meixner_1992}%
  \BibitemOpen
  \bibfield  {author} {\bibinfo {author} {\bibfnamefont {A.~J.}\ \bibnamefont
  {Meixner}}, \bibinfo {author} {\bibfnamefont {C.~M.}\ \bibnamefont
  {Jefferson}}, \ and\ \bibinfo {author} {\bibfnamefont {R.~M.}\ \bibnamefont
  {Macfarlane}},\ }\href {\doibase 10.1103/PhysRevB.46.5912} {\bibfield
  {journal} {\bibinfo  {journal} {Physical Review B}\ }\textbf {\bibinfo
  {volume} {46}},\ \bibinfo {pages} {5912} (\bibinfo {year}
  {1992})}\BibitemShut {NoStop}%
\bibitem [{\citenamefont {Li}\ \emph {et~al.}(2016)\citenamefont {Li},
  \citenamefont {Bao}, \citenamefont {Thuresson}, \citenamefont {Nilsson},
  \citenamefont {Rippe},\ and\ \citenamefont {Kröll}}]{li_2016}%
  \BibitemOpen
  \bibfield  {author} {\bibinfo {author} {\bibfnamefont {Q.}~\bibnamefont
  {Li}}, \bibinfo {author} {\bibfnamefont {Y.}~\bibnamefont {Bao}}, \bibinfo
  {author} {\bibfnamefont {A.}~\bibnamefont {Thuresson}}, \bibinfo {author}
  {\bibfnamefont {A.~N.}\ \bibnamefont {Nilsson}}, \bibinfo {author}
  {\bibfnamefont {L.}~\bibnamefont {Rippe}}, \ and\ \bibinfo {author}
  {\bibfnamefont {S.}~\bibnamefont {Kröll}},\ }\href {\doibase
  10.1103/PhysRevA.93.043832} {\bibfield  {journal} {\bibinfo  {journal}
  {Physical Review A}\ }\textbf {\bibinfo {volume} {93}},\ \bibinfo {pages}
  {043832} (\bibinfo {year} {2016})}\BibitemShut {NoStop}%
\bibitem [{\citenamefont {Amari}\ \emph {et~al.}(2010)\citenamefont {Amari},
  \citenamefont {Walther}, \citenamefont {Sabooni}, \citenamefont {Huang},
  \citenamefont {Kröll}, \citenamefont {Afzelius}, \citenamefont {Usmani},
  \citenamefont {Lauritzen}, \citenamefont {Sangouard}, \citenamefont
  {de~Riedmatten},\ and\ \citenamefont {Gisin}}]{amari_2010}%
  \BibitemOpen
  \bibfield  {author} {\bibinfo {author} {\bibfnamefont {A.}~\bibnamefont
  {Amari}}, \bibinfo {author} {\bibfnamefont {A.}~\bibnamefont {Walther}},
  \bibinfo {author} {\bibfnamefont {M.}~\bibnamefont {Sabooni}}, \bibinfo
  {author} {\bibfnamefont {M.}~\bibnamefont {Huang}}, \bibinfo {author}
  {\bibfnamefont {S.}~\bibnamefont {Kröll}}, \bibinfo {author} {\bibfnamefont
  {M.}~\bibnamefont {Afzelius}}, \bibinfo {author} {\bibfnamefont
  {I.}~\bibnamefont {Usmani}}, \bibinfo {author} {\bibfnamefont
  {B.}~\bibnamefont {Lauritzen}}, \bibinfo {author} {\bibfnamefont
  {N.}~\bibnamefont {Sangouard}}, \bibinfo {author} {\bibfnamefont
  {H.}~\bibnamefont {de~Riedmatten}}, \ and\ \bibinfo {author} {\bibfnamefont
  {N.}~\bibnamefont {Gisin}},\ }\href {\doibase 10.1016/j.jlumin.2010.01.012}
  {\bibfield  {journal} {\bibinfo  {journal} {Journal of Luminescence}\
  }\textbf {\bibinfo {volume} {130}},\ \bibinfo {pages} {1579} (\bibinfo {year}
  {2010})}\BibitemShut {NoStop}%
\bibitem [{\citenamefont {Equall}\ \emph {et~al.}(1995)\citenamefont {Equall},
  \citenamefont {Cone},\ and\ \citenamefont {Macfarlane}}]{equall_1995}%
  \BibitemOpen
  \bibfield  {author} {\bibinfo {author} {\bibfnamefont {R.~W.}\ \bibnamefont
  {Equall}}, \bibinfo {author} {\bibfnamefont {R.~L.}\ \bibnamefont {Cone}}, \
  and\ \bibinfo {author} {\bibfnamefont {R.~M.}\ \bibnamefont {Macfarlane}},\
  }\href {\doibase 10.1103/PhysRevB.52.3963} {\bibfield  {journal} {\bibinfo
  {journal} {Physical Review B}\ }\textbf {\bibinfo {volume} {52}},\ \bibinfo
  {pages} {3963} (\bibinfo {year} {1995})}\BibitemShut {NoStop}%
\bibitem [{\citenamefont {Könz}\ \emph {et~al.}(2003)\citenamefont {Könz},
  \citenamefont {Sun}, \citenamefont {Thiel}, \citenamefont {Cone},
  \citenamefont {Equall}, \citenamefont {Hutcheson},\ and\ \citenamefont
  {Macfarlane}}]{konz2003}%
  \BibitemOpen
  \bibfield  {author} {\bibinfo {author} {\bibfnamefont {F.}~\bibnamefont
  {Könz}}, \bibinfo {author} {\bibfnamefont {Y.}~\bibnamefont {Sun}}, \bibinfo
  {author} {\bibfnamefont {C.~W.}\ \bibnamefont {Thiel}}, \bibinfo {author}
  {\bibfnamefont {R.~L.}\ \bibnamefont {Cone}}, \bibinfo {author}
  {\bibfnamefont {R.~W.}\ \bibnamefont {Equall}}, \bibinfo {author}
  {\bibfnamefont {R.~L.}\ \bibnamefont {Hutcheson}}, \ and\ \bibinfo {author}
  {\bibfnamefont {R.~M.}\ \bibnamefont {Macfarlane}},\ }\href {\doibase
  10.1103/PhysRevB.68.085109} {\bibfield  {journal} {\bibinfo  {journal}
  {Physical Review B}\ }\textbf {\bibinfo {volume} {68}},\ \bibinfo {pages}
  {085109} (\bibinfo {year} {2003})}\BibitemShut {NoStop}%
\bibitem [{\citenamefont {Thorpe}\ \emph {et~al.}(2011)\citenamefont {Thorpe},
  \citenamefont {Rippe}, \citenamefont {Fortier}, \citenamefont {Kirchner},\
  and\ \citenamefont {Rosenband}}]{thorpe2011}%
  \BibitemOpen
  \bibfield  {author} {\bibinfo {author} {\bibfnamefont {M.~J.}\ \bibnamefont
  {Thorpe}}, \bibinfo {author} {\bibfnamefont {L.}~\bibnamefont {Rippe}},
  \bibinfo {author} {\bibfnamefont {T.~M.}\ \bibnamefont {Fortier}}, \bibinfo
  {author} {\bibfnamefont {M.~S.}\ \bibnamefont {Kirchner}}, \ and\ \bibinfo
  {author} {\bibfnamefont {T.}~\bibnamefont {Rosenband}},\ }\href {\doibase
  10.1038/nphoton.2011.215} {\bibfield  {journal} {\bibinfo  {journal} {Nature
  Photonics}\ }\textbf {\bibinfo {volume} {5}},\ \bibinfo {pages} {688}
  (\bibinfo {year} {2011})}\BibitemShut {NoStop}%
\bibitem [{\citenamefont {Leibrandt}\ \emph {et~al.}(2013)\citenamefont
  {Leibrandt}, \citenamefont {Thorpe}, \citenamefont {Chou}, \citenamefont
  {Fortier}, \citenamefont {Diddams},\ and\ \citenamefont
  {Rosenband}}]{leibrandt2013}%
  \BibitemOpen
  \bibfield  {author} {\bibinfo {author} {\bibfnamefont {D.~R.}\ \bibnamefont
  {Leibrandt}}, \bibinfo {author} {\bibfnamefont {M.~J.}\ \bibnamefont
  {Thorpe}}, \bibinfo {author} {\bibfnamefont {C.-W.}\ \bibnamefont {Chou}},
  \bibinfo {author} {\bibfnamefont {T.~M.}\ \bibnamefont {Fortier}}, \bibinfo
  {author} {\bibfnamefont {S.~A.}\ \bibnamefont {Diddams}}, \ and\ \bibinfo
  {author} {\bibfnamefont {T.}~\bibnamefont {Rosenband}},\ }\href {\doibase
  10.1103/PhysRevLett.111.237402} {\bibfield  {journal} {\bibinfo  {journal}
  {Physical Review Letters}\ }\textbf {\bibinfo {volume} {111}},\ \bibinfo
  {pages} {237402} (\bibinfo {year} {2013})}\BibitemShut {NoStop}%
\bibitem [{\citenamefont {Gobron}\ \emph {et~al.}(2017)\citenamefont {Gobron},
  \citenamefont {Jung}, \citenamefont {Galland}, \citenamefont {Predehl},
  \citenamefont {Targat}, \citenamefont {Ferrier}, \citenamefont {Goldner},
  \citenamefont {Seidelin},\ and\ \citenamefont {Coq}}]{gobron2017}%
  \BibitemOpen
  \bibfield  {author} {\bibinfo {author} {\bibfnamefont {O.}~\bibnamefont
  {Gobron}}, \bibinfo {author} {\bibfnamefont {K.}~\bibnamefont {Jung}},
  \bibinfo {author} {\bibfnamefont {N.}~\bibnamefont {Galland}}, \bibinfo
  {author} {\bibfnamefont {K.}~\bibnamefont {Predehl}}, \bibinfo {author}
  {\bibfnamefont {R.~L.}\ \bibnamefont {Targat}}, \bibinfo {author}
  {\bibfnamefont {A.}~\bibnamefont {Ferrier}}, \bibinfo {author} {\bibfnamefont
  {P.}~\bibnamefont {Goldner}}, \bibinfo {author} {\bibfnamefont
  {S.}~\bibnamefont {Seidelin}}, \ and\ \bibinfo {author} {\bibfnamefont
  {Y.~L.}\ \bibnamefont {Coq}},\ }\href {\doibase 10.1364/OE.25.015539}
  {\bibfield  {journal} {\bibinfo  {journal} {Optics Express}\ }\textbf
  {\bibinfo {volume} {25}},\ \bibinfo {pages} {15539} (\bibinfo {year}
  {2017})}\BibitemShut {NoStop}%
\bibitem [{\citenamefont {Ma}\ \emph {et~al.}(2012)\citenamefont {Ma},
  \citenamefont {Slattery},\ and\ \citenamefont {Tang}}]{Lijun_upconversion}%
  \BibitemOpen
  \bibfield  {author} {\bibinfo {author} {\bibfnamefont {L.}~\bibnamefont
  {Ma}}, \bibinfo {author} {\bibfnamefont {O.}~\bibnamefont {Slattery}}, \ and\
  \bibinfo {author} {\bibfnamefont {X.}~\bibnamefont {Tang}},\ }\href {\doibase
  10.1016/j.physrep.2012.07.006} {\bibfield  {journal} {\bibinfo  {journal}
  {Physics Reports}\ }\textbf {\bibinfo {volume} {521}},\ \bibinfo {pages} {69
  } (\bibinfo {year} {2012})}\BibitemShut {NoStop}%
\bibitem [{\citenamefont
  {Ran{\ifmmode\check{c}\else\v{c}\fi}i{\ifmmode\acute{c}\else\'{c}\fi}}\ \emph
  {et~al.}(2018)\citenamefont
  {Ran{\ifmmode\check{c}\else\v{c}\fi}i{\ifmmode\acute{c}\else\'{c}\fi}},
  \citenamefont {Hedges}, \citenamefont {Ahlefeldt},\ and\ \citenamefont
  {Sellars}}]{Rancic_2018}%
  \BibitemOpen
  \bibfield  {author} {\bibinfo {author} {\bibfnamefont {M.}~\bibnamefont
  {Ran{\ifmmode\check{c}\else\v{c}\fi}i{\ifmmode\acute{c}\else\'{c}\fi}}},
  \bibinfo {author} {\bibfnamefont {M.~P.}\ \bibnamefont {Hedges}}, \bibinfo
  {author} {\bibfnamefont {R.~L.}\ \bibnamefont {Ahlefeldt}}, \ and\ \bibinfo
  {author} {\bibfnamefont {M.~J.}\ \bibnamefont {Sellars}},\ }\href {\doibase
  10.1038/nphys4254} {\bibfield  {journal} {\bibinfo  {journal} {Nature
  Physics}\ }\textbf {\bibinfo {volume} {14}},\ \bibinfo {pages} {50} (\bibinfo
  {year} {2018})}\BibitemShut {NoStop}%
\bibitem [{\citenamefont {Böttger}\ \emph {et~al.}(2009)\citenamefont
  {Böttger}, \citenamefont {Thiel}, \citenamefont {Cone},\ and\ \citenamefont
  {Sun}}]{bottger_2009}%
  \BibitemOpen
  \bibfield  {author} {\bibinfo {author} {\bibfnamefont {T.}~\bibnamefont
  {Böttger}}, \bibinfo {author} {\bibfnamefont {C.~W.}\ \bibnamefont {Thiel}},
  \bibinfo {author} {\bibfnamefont {R.~L.}\ \bibnamefont {Cone}}, \ and\
  \bibinfo {author} {\bibfnamefont {Y.}~\bibnamefont {Sun}},\ }\href {\doibase
  10.1103/PhysRevB.79.115104} {\bibfield  {journal} {\bibinfo  {journal}
  {Physical Review B}\ }\textbf {\bibinfo {volume} {79}},\ \bibinfo {pages}
  {115104} (\bibinfo {year} {2009})}\BibitemShut {NoStop}%
\bibitem [{\citenamefont {Böttger}\ \emph {et~al.}(2006)\citenamefont
  {Böttger}, \citenamefont {Sun}, \citenamefont {Thiel},\ and\ \citenamefont
  {Cone}}]{bottger_2006}%
  \BibitemOpen
  \bibfield  {author} {\bibinfo {author} {\bibfnamefont {T.}~\bibnamefont
  {Böttger}}, \bibinfo {author} {\bibfnamefont {Y.}~\bibnamefont {Sun}},
  \bibinfo {author} {\bibfnamefont {C.~W.}\ \bibnamefont {Thiel}}, \ and\
  \bibinfo {author} {\bibfnamefont {R.~L.}\ \bibnamefont {Cone}},\ }\href
  {\doibase 10.1103/PhysRevB.74.075107} {\bibfield  {journal} {\bibinfo
  {journal} {Physical Review B}\ }\textbf {\bibinfo {volume} {74}},\ \bibinfo
  {pages} {075107} (\bibinfo {year} {2006})}\BibitemShut {NoStop}%
\bibitem [{\citenamefont {Rippe}\ \emph {et~al.}(2005)\citenamefont {Rippe},
  \citenamefont {Nilsson}, \citenamefont {Kröll}, \citenamefont {Klieber},\
  and\ \citenamefont {Suter}}]{rippe_2005}%
  \BibitemOpen
  \bibfield  {author} {\bibinfo {author} {\bibfnamefont {L.}~\bibnamefont
  {Rippe}}, \bibinfo {author} {\bibfnamefont {M.}~\bibnamefont {Nilsson}},
  \bibinfo {author} {\bibfnamefont {S.}~\bibnamefont {Kröll}}, \bibinfo
  {author} {\bibfnamefont {R.}~\bibnamefont {Klieber}}, \ and\ \bibinfo
  {author} {\bibfnamefont {D.}~\bibnamefont {Suter}},\ }\href {\doibase
  10.1103/PhysRevA.71.062328} {\bibfield  {journal} {\bibinfo  {journal}
  {Physical Review A}\ }\textbf {\bibinfo {volume} {71}},\ \bibinfo {pages}
  {062328} (\bibinfo {year} {2005})}\BibitemShut {NoStop}%
\bibitem [{\citenamefont {Roos}\ and\ \citenamefont
  {Mølmer}(2004)}]{roos_2004}%
  \BibitemOpen
  \bibfield  {author} {\bibinfo {author} {\bibfnamefont {I.}~\bibnamefont
  {Roos}}\ and\ \bibinfo {author} {\bibfnamefont {K.}~\bibnamefont {Mølmer}},\
  }\href {\doibase 10.1103/PhysRevA.69.022321} {\bibfield  {journal} {\bibinfo
  {journal} {Physical Review A}\ }\textbf {\bibinfo {volume} {69}},\ \bibinfo
  {pages} {022321} (\bibinfo {year} {2004})}\BibitemShut {NoStop}%
\bibitem [{\citenamefont {Sun}(2006)}]{sun_rare_2006}%
  \BibitemOpen
  \bibfield  {author} {\bibinfo {author} {\bibfnamefont {Y.~C.}\ \bibnamefont
  {Sun}},\ }in\ \href@noop {} {\emph {\bibinfo {booktitle} {Spectroscopic
  {Properties} of {Rare} {Earths} in {Optical} {Materials}}}},\ \bibinfo
  {editor} {edited by\ \bibinfo {editor} {\bibfnamefont {G.}~\bibnamefont
  {Liu}}\ and\ \bibinfo {editor} {\bibfnamefont {B.}~\bibnamefont {Jacquier}}}\
  (\bibinfo  {publisher} {Springer Science \& Business Media},\ \bibinfo {year}
  {2006})\ pp.\ \bibinfo {pages} {379--429}\BibitemShut {NoStop}%
\bibitem [{\citenamefont {Chang}\ \emph {et~al.}(2005)\citenamefont {Chang},
  \citenamefont {Tian}, \citenamefont {Mohan}, \citenamefont {Renner},
  \citenamefont {Merkel},\ and\ \citenamefont {Babbitt}}]{chang_2005}%
  \BibitemOpen
  \bibfield  {author} {\bibinfo {author} {\bibfnamefont {T.}~\bibnamefont
  {Chang}}, \bibinfo {author} {\bibfnamefont {M.}~\bibnamefont {Tian}},
  \bibinfo {author} {\bibfnamefont {R.~K.}\ \bibnamefont {Mohan}}, \bibinfo
  {author} {\bibfnamefont {C.}~\bibnamefont {Renner}}, \bibinfo {author}
  {\bibfnamefont {K.~D.}\ \bibnamefont {Merkel}}, \ and\ \bibinfo {author}
  {\bibfnamefont {W.~R.}\ \bibnamefont {Babbitt}},\ }\href {\doibase
  10.1364/OL.30.001129} {\bibfield  {journal} {\bibinfo  {journal} {Optics
  Letters}\ }\textbf {\bibinfo {volume} {30}},\ \bibinfo {pages} {1129}
  (\bibinfo {year} {2005})}\BibitemShut {NoStop}%
\end{thebibliography}
%

\end{document}